\documentstyle[12pt,sec]{article}
\makeatletter
\@addtoreset{equation}{section}
\makeatother

\def\lsim{\mathrel{\rlap{\lower4pt\hbox{\hskip1pt$\sim$}}
    \raise1pt\hbox{$<$}}}         
\def\gsim{\mathrel{\rlap{\lower4pt\hbox{\hskip1pt$\sim$}}
    \raise1pt\hbox{$>$}}}         

\def\ut#1{$\underline{\smash{\vphantom{y}\hbox{#1}}}$}
\def\overleftrightarrow#1{\vbox{\ialign{##\crcr
    $\leftrightarrow$\crcr
    \noalign{\kern 1pt\nointerlineskip}
    $\hfil\displaystyle{#1}\hfil$\crcr}}}
\long\def\caption#1#2{{\setbox1=\hbox{#1\quad}\hbox{\copy1%
\vtop{\advance\hsize by -\wd1 \noindent #2}}}}
\newcommand{\asmu}{\alpha_s(\mu^2)}
\newcommand \beq{\begin{eqnarray}}
\newcommand \eeq{\end{eqnarray}}

\newcommand \la{\raisebox{-.5ex}{$\stackrel{<}{\sim}$}}

\newcommand \ps {P(\sigma)}
\newcommand \s{\bar{\sigma}}

\newcommand \os{\omega_\sigma}

\newcommand{\av}[1]{\langle{#1}\rangle}
\newcommand{\rep}[1]{(\ref{#1})}
\newcommand{\bea}{\begin{eqnarray}}
\newcommand{\eea}{\end{eqnarray}}
\newcommand{\ket}[1]{|{#1}\rangle}

\newcommand{\half}{{\scriptstyle \frac12 }}

\newcommand \sd{\sigma_{diff}}

\begin{document}
\bibliographystyle{unsrt}
\baselineskip=20pt
\centerline{\bf  The Geometrical Color Optics Of Coherent High Energy
Processes }
\vskip 6pt
\centerline{L.L.Frankfurt}
\vskip 6pt
\centerline{\it Dep't of Physics, Tel Aviv University,69978 Ramat Aviv, Israel}
\centerline {\it leave of absence of Inst. of Nuclear Physics, }
\centerline{\it  St. Petersburg, Russia.}
\vskip 6pt
\centerline{G.A. Miller}
\vskip 6pt
\centerline{\it Nuclear Theory Group, Department of Physics, FM-15}
\centerline{\it University of Washington, Seattle, Washington 98195}
\vskip 6pt
\centerline{and}
\vskip 6pt
\centerline{M. Strikman}
\vskip 6pt
\centerline{\it Department of Physics}
\centerline{\it Pennsylvania State University, University Park, PA 16802}
\centerline {\it and Inst. of Nuclear Physics, St. Petersburg, Russia.}
\vskip 6pt

\date{ }


\vfill\eject
\noindent   \ut{Contents}
\vskip 12pt

\noindent 1.Introduction.

1.1 QCD ideas to be tested

\noindent 2.Coherence phenomena in QED

2.1 Charge transparency

2.2 Charge filtering

2.3 Charge opacity

\noindent 3. Color transparency in perturbative QCD

3.1 Coherence length in QCD

3.2 Bjorken scaling for deep inelastic processes

3.3 Hard Coherent Diffraction From Nucleons and Nuclei

\noindent 4. Soft Diffractive Physics and Color Fluctuations

4.1 Scattering eigenstates formalism

4.2 Current information on color fluctuations in Hadrons

4.3 Small Size Configurations in Pions

4.4 Nuclear inelastic  diffraction as evidence for
 color fluctuations near the average value

4.5  Inelastic screening corrections to the total cross sections

\noindent 5. Color transparency in nuclear quasielastic reactions

5.1 Is a small system made?

5.2 Time development

5.3 Relevant data

\noindent 6. Color fluctuations in nucleons and nucleus-nucleus collisions

6.1 Color opacity effects

6.2 Color fluctuations and central collisions

6.3 Leading hadron transparency effects in central collisions

\noindent 7. Outlook

\section{Introduction}
\label{sec:intro}

     The aim of this review is to explain the
expansion of quantum
chromodynamics QCD to the domain of
coherent phenomena in high energy (hundreds and thousands of GeV)
and in medium energy (several GeV) processes,
including the realization that some hard diffractive processes
and related color transparency phenomenon
can be  calculated using QCD in a model independent way.
It seems now
that exploring color coherent
phenomena may produce new
effective methods to investigate  QCD,  hadron and  nuclear
structure, and even help disentangle
and vary the properties of
hadronic systems detected in heavy ion  collisions.

Although
twenty years of experimental investigations of high momentum transfer
``hard"
processes have confirmed the
basic predictions of perturbative $QCD$\cite{altarelli,fwarnps},
 confinement of quarks and gluons, as well as
the  physics of spontaneously
broken chiral symmetry, are not
yet understood.
The study of coherent phenomena in which
color aspects of QCD as a quantum field theory
play a dominant role provides new
experimentally testable ideas. For example, suppose a high momentum,
color singlet (color neutral) system
of closely separated
quarks and gluons is produced in a collision
and  emerges as a detected particle. Suppose further that
the system interacts by radiating long wavelength
gluons which are absorbed by a  target.
Here gluon
radiation effects are computed by adding the contributions from each of the
constituents and then squaring the matrix elements, i.e. the process is
coherent. But the leading term,  in which one simply adds up the color
charges, vanishes because of the color neutrality of the system.
Thus the ``color monopole" term vanishes and one next looks at the
effects of color dipole terms. However, if the system is small enough
the dipole effects are small,
and interactions with the target are suppressed, or ``color screened".

The color screening process depicted here is analogous
to some features of electrically neutral systems in quantum
electrodynamics QED.
For example, $\mu^+-\mu^-$ ground state is smaller
than positronium and is therefore expected to have weaker interactions with
media.  But the reduction of interactions for small-sized objects
is a rare property.
There are many other theories,
such as the sigma model of bare nucleons and mesons,
 in which there is no concept
of neutrality, and no expected
decrease of interaction strength with
size. The similarities between
QCD  and the well-understood
QED allows us to use analogies to clarify various  points.
But we shall also remember
that QCD is a non-Abelian theory (gluons
interact with each other), and shall discuss some effects
not inherent in QED.

\subsection{QCD ideas to be tested}

QCD has  important implications for high energy processes:
i) small objects have reduced interactions;
ii) hadrons consist
of
 configurations of very different spatial sizes;
iii) at sufficiently high energies,
 hadronic quark-gluon configurations
can be considered as frozen
because the coherence length controlling whether
wave mechanics
or geometrical color optics is to be used increases with energy;
iv) small objects are produced at high $Q^2$ in two-body
high momentum transfer Q$^2$ reactions;
v) at insufficiently high energies, a small object will expand;
vi) large sized objects can be captured in central heavy ion collisions.

The notions i)
ii) and iii) have diverse consequences.
One is
color transparency. Suppose a small-sized object is made. This doesn't
interact much according to color screening, and
at high enough energies, this
object can pass through an
entire nucleus without fluctuating into a configuration of larger size.
Thus it doesn't interact, and the nucleus is transparent.
Color transparency can appear
in a variety of processes:   diffractive
excitation of pions \cite{Bertsch81}
especially into two jets \cite{FMS93},
 hard nuclear diffractive processes
\cite{Fra81}
,in small cross sections for $J\Psi$ meson photoproduction
\cite{FK},
 leading
hadron production in
 nuclear reactions using $\mu^+-\mu^-$ (Drell-Yan)
pairs as a trigger \cite{FS85},
and the electroproduction of vector mesons \cite{FS88,BM88}.
However, a  recent development in QCD is that
to completely eliminate all interactions
(as opposed to suppress)
it is necessary to increase
both the energy and decrease the size of
the configuration, Eq.~\rep{sigmaqqbar}.
If the size of configuration is fixed
 the interaction of a small object
is more shadowed as the  energy increases.
\cite{FMS93,Brod94}. This
reflects the important role of soft, nonperturbative
physics in the limit of fixed $Q^2$ and high energy.

If iv) holds then color transparency can be observed in
high Q$^2$ nuclear quasielastic scattering\cite{Mue82,Bro82}.
But the present experiments\cite{bnl88,eva,ne18}
are at energies too low
for the validity of the frozen approximation ii).
Thus, v) the effects of quantum fluctuations of a small-sized configuration
into one of a larger
size  must be taken into account.
Thus  nuclear quasielastic processes provide  unique
possibilities for laboratory observations of the
restoration of long wavelength (soft)
color, pion....
fields
\cite{FLFS89,FMS92}.
  Nucleus-nucleus collisions provide an
arena to observe the  effects of color transparency such as in the
production of leading nucleons\cite{FS81,FS91a,FS91}. But the complementary
color opacity effect vi) in which large-sized configurations are absorbed
strongly can lead to
significant fluctuations of transverse energy for central collisions
\cite{HBB91,BBF91} , and to
transitions to percolated configurations
\cite{GB78}, where ``fat"
nucleons form a connected net in the nucleus \cite{BF93}.

\subsubsection {Small objects have reduced interactions - color screening}

 Low\cite{Low75},
and Gunion and Soper\cite{gs77},
used a   two-gluon exchange model\cite{Low75,Nus75} to
observe that
in QCD  the
interaction between a small sized hadronic system and a target
is largely
suppressed. This suppression,
called color screening, is one of the essential ingredients needed to
understand how
Bjorken scaling for  deep inelastic scattering at small x
occurs.

The color screening
explanation of Bjorken scaling
at small x
was discussed in Refs.~\cite{FS88,FS91,FS89,nnn}
But the two-gluon exchange model is not complete.
At small x, QCD demands that a small size configuration
interacts with a target by producing a
soft self-interacting (non-Abelian)
multi-gluon field so that two gluon exchange diagrams
do not adequately describe the process. Instead,
the renormalizability, gauge invariance
 and asymptotic
freedom
of QCD  can be used to prove that, the
interaction, maintains color screening but is proportional
 for  some kinematics
to the target
gluon distribution
as in Eq. \rep{sigmaqqbar}
\cite {FS89hi,BBFS93}.

ii)   {\bf Configurations of very different sizes exist in hadrons}.

It is expected, and lattice calculations indicate
that in QCD the
so-called elementary particles are
bound states of strongly interacting quarks and gluons.
 The hadron can be described in terms of an infinite number of
basis states, or configurations,
for example $|qqq>$, $|qqqg>$, etc.
Such different configurations are expected to have
varying sizes. In quantum mechanics, a system fluctuates between its
different configurations.
Thus ``snapshots" of a hadron taken
at different times would observe both small and
large sized configurations; we call this color fluctuations.
In the hadronic rest frame, the uncertainty principle estimates the time
scale for fluctuations between two
configurations the inverse of the energy or mass difference
of the two configurations, $\tau\sim \hbar/ (m -M)$. The relevant mass
differences are
typically of the order of hundreds of MeV,
so the fluctuation time is very small about 1 fm/c.

iii){\bf At sufficiently
 high energies hadronic quark-gluon configurations
can be considered as frozen}.
Suppose the hadron has a large laboratory momentum $p_{lab}$, with
$v\approx c=1$.
The energy difference between configurations of mass m and M
becomes $\sqrt{p_{lab}^2 + M^2}-\sqrt{p_{lab}^2 + m^2}$.
 If $p_{lab}$ is very large,
the energy denominator $\approx (m^2-M^2)/2 p_{lab}$ can be small,
the time scale for fluctuation is long, and
the configuration of mass M can move for a long
distance, $l_c$
before decaying. Thus we have
\bea
l_c=2p_{lab}/(M^2-m^2).\label{eq:lc}
\eea
Coherent, constructively interfering
interactions occur between the excited configuration and
target material spread over the length $l_c$, hence the name coherence
length.
Such a coherence length exists in any quantum field theory.
   Landau and Pomeranchuk \cite {LP}
 were the first who found this coherence in
$QED$.
The existence of  coherence length in QED leads
to a strong suppression of soft photon
radiation accompanying energetic electrons  moving through  a medium;
see Refs.~\cite{LP,Migdal,TerMik,Feinberg} and references therein.
  Later on it was understood that the coherence length exists
in  strong interactions as well
\cite{Gribov65,Gribov69,gribovschool,FS88}.
The  long time required for transitions between configurations of
a high momentum hadron
is the basis of Feynman's parton model.
Note that $l_c$ grows with energy and can be
greater than the radius of a  target nucleus, then   the
configuration can pass through without decaying \cite{Mandelstam}.

The  coherence length is an important feature of
 deep inelastic scattering \cite{Gribov65}
and diffractive dissociation of hadrons\cite{FP56}.
In deep inelastic scattering, at low x,
the time or length
scale for a photon to fluctuate into a  hadronic or $q \bar q$ component
is given by
$l_c=1/(2 M_N x) \approx (0.1/x) {\rm fm}, $
where $M_N$ is the nucleon mass, x= $Q^2/(2 M_N \nu)$ where $\nu$ is the
energy of the virtual photon. At sufficiently low x, $l_c$ can be much
larger than the nuclear diameter so that  the fluctuation can
occur long before the $q \bar q$ pair hits the target.
In diffractive dissociation an incident high energy particle of mass m
can be excited into a state of mass M,
while leaving the target in its ground state.
 The longitudinal momentum
transfer $q_\parallel\approx (M^2-m^2)/2 p_{lab}$ so that
contributions of target matter spread
over a very large longitudinal distances
$1/q_\parallel\approx \l_c$ interfere constructively.

Another consequence of a large value of $l_c$ is that an
incident hadron, in a specific configuration,
stays in that configuration as it passes through the entire target.
 Then it is possible to describe
high energy diffractive processes in terms
of a probability $P(\sigma)$ that a
configuration interacts with a cross section
$\sigma$\cite {FP56,GW60}.  $P(\sigma)$, when
reconstructed from data
 is quite broad,  Sect. \rep{subs:info},  so that
different configurations interact with widely varying strengths
as noted in Fig.~1.
This gives an opportunity to observe both color transparency and color
opacity effects.
Note that we neglect the change
of
the  impact parameter distribution
in the Pomeron  ladder  at high energies
(``Gribov diffusion") \cite{gribovschool}. This
plays an important role at collider energies.

iv) {\bf Small objects are produced at high $Q^2$ in two-body
wide angle reactions.}
High momentum  transfer is usually associated with small
wavelengths. But iv) asserts more, that an object of small size,
  without soft color and pion fields,
 is produced during
a high momentum transfer process. This
was suggested originally
as a consequence of perturbative QCD applied at very high Q$^2$
\cite{Mue82,Bro82}.
 But the relevance of perturbative QCD for experimentally accessible
values of Q$^2$
was questioned by
several authors \cite{ar,isgurls}. More recently Sterman and collaborators
\cite{Botts,sterman}
showed how including Sudakov
type
effects \cite{sudakov} extended the region
where perturbative calculations could be applied.
However, it is relevant to ask if small sized objects are produced
by non-perturbative  effects.
Indeed,
theoretical analyses of the simplest versions of popular hadronic models show
 \cite{FMS92,FMS931} that "small configurations"  can be produced
for momentum transfers as low as about 1-2 (GeV/c)$^2$.

v){\bf At insufficiently high energies a point-like configuration will
expand.}  For sufficiently hard processes  perturbative QCD unambiguously
predicts that a very small or point like configuration PLC will be formed
and that color transparency will occur. However at moderate $Q^2$   a new
interesting phenomenon  is important.  A point-like configuration is a
coherent superposition of eigenstates of the QCD Hamiltonian which is  of
essentially no size. Such a  wave packet undergoes time evolution which can
only increase its size
and restore soft quark-gluon fields. %
Consider the time it takes for a small object to expand to a normal
hadronic size of about 1 fm. The time $\tau$ for a quantum fluctuation from
a point-like configuration
(of bare mass M) to fluctuate into an object of normal hadronic size is
given by  $\tau= 2 p_{lab}/(M^2 -  m^2)$, with m as the ground state mass.
Thus the expansion time is controlled by an undetermined parameter M. For
sufficiently large energies, $\tau$ is long enough so that the object can
leave the nucleus while small enough to avoid final state interactions, and
color transparency occurs. But what is M? Purely theoretical arguments do
not give a value, but a lower limit is the sum of the nucleon and pion
masses. Thus $\tau\le (E/1 GeV) fm$,  which for the experimentally relevant
value of E $\approx $ 5 GeV,  is about 5 fm or smaller than the diameter of
Al. Thus for present experiments, the $PLC$ must expand significantly as it
moves through the nucleus. Different estimates based on the quark model,
Regge model, and the resonance spectra of pp diffractive processes
necessarily lead to smaller values of $\tau$. Thus expansion must occur and
final state interactions are not completely suppressed. However, the
expansion rate is not yet determined by experiment, so existing
calculations harbor uncertainties.

The point-like
configuration is very small and can therefore be thought of as a small
bit of matter in the perturbative phase, a bubble,  moving within
the nonperturbative QCD medium. Space-time evolution, which
increases the size of the PLC, does so by
a  restoration of soft fields.
The time development  of a bubble
of one phase within another
phase  appears to be  one of the key elements in the
construction of the
theory of heavy ion collisions \cite{RW92}  and theory of the early Universe.
Laboratory investigations of this physics
can be done with
nuclear  quasi-elastic reactions which search for color transparency.
 The BNL (p,pp)
data\cite{bnl88}
and the NE18 (e,e'p) data\cite{ne18}
are consistent with
the notion that a PLC is made and expands before leaving the nucleus.

vi){\bf Large sized objects can be captured in central heavy ion collision}

An incident nucleon in a large
sized configuration should
have much stronger interactions than a nucleon in an
average size collision.
Consider now a head-on low impact parameter collision
between two heavy ions. This is called a central
collision and many particles are produced.
Large sized configurations interact more strongly
 and produce more particles than other
configurations. Hence central collisions are strongly influenced by large
sized configurations- such configurations stand out and
can be said to have been produced.
There are many possible consequences of the notion that large sized
configurations can be important.
For example,
Heiselberg et al  \cite{HBB91,BBF91}  have argued that
color fluctuations
can account for the large fluctuations experimentally observed in
$E_t$ ($\approx p_t$) distributions, and also
may help to
regulate parton distributions of colliding
nuclei \cite{BF93}.

\subsection{Outline}

Sect.~\ref{sec:qed} is concerned with using QED to provide simple
examples of charge
transparency, opacity and filtering.
  For example, the influence of
charge screening on a sufficiently
energetic small sized $e^+ e^-$ pair
leads to an
experimentally  observed charge transparency
\cite{Chud55,Perc55,ziel}.
This and related effects of charge filtering and charge
opacity are discussed.

Sect.~\ref{sec:pqcd} is concerned with
color screening effects in the
high energy domain. Deep inelastic scattering
at low x, is reviewed first.
The vanishing of the
q$\bar q$-target interactions for small transverse
separations is required for scaling to occur.
It has been shown
recently\cite{BBFS93}
that such interactions depend upon the gluon distribution $G_T$ of the
target
and therefore
they are screened.
The amount of
nuclear shadowing in
$G_T(x,Q^2)$ depends on
Q$^2$ it is governed by
the evolution equations of QCD, and  consequences
\cite{mq86,fsl,BBFS93,FMS93,Brod94}.
of this are discussed.

Then
in Sect.
\ref{subs:hcd} we explain that a number of
hard
 diffractive processes involving nucleon and nuclear targets
can be legitimately calculated in QCD.
Cross  sections for hard coherent
electroproduction of vector mesons;
coherent and incoherent diffraction of pions into two
jets:  $\pi +A \rightarrow 2 jets + A (A^*) $ have been
calculated in
QCD
and are proportional to the square of
wave functions of the minimal Fock components
of
hadronic and nuclear light-cone wave functions \cite {FMS93,Brod94}.
The results are in
agreement with the available EMC, NMC and E-665 FNAL  data
 for $\rho$
electroproduction.
  Thus experiments at FNAL and HERA could study
light-cone wave functions of hadrons and nuclei, including
gluon distributions. Calculating such wave functions remains an
interesting challenge
for the
nonperturbative  lattice and dispersion sum rule methods.
Specific quantum field theory effects are important here,
in that
QCD and
phenomenological quark models predict
different atomic number
dependences.
In particular,  in Ref.~\cite {KNNZ93} nuclear shadowing is a higher twist
effect, in contrast with the leading twist result of QCD for
the production of longitudinally and transversely polarized
vector mesons  which specifies even more shadowing for the
transverse case\cite{Brod94}.
But the application of
the nonrelativistic quark model to hard
processes is questionable
here, since in QCD, amplitudes are expressed in terms of
distributions of current, not constituent quarks(\cite{BL80}
and refs therein).

The consequences
of color fluctuations are discussed in
Sec.~\rep{sec:soft}.
One result is that  there is significant  probability for a nucleon
 (pion) to be in a configuration with small $\sigma$, e.g.
 $P_N(\sigma \le 5 mb) \approx  2\%$
for the pion.
Small values of $\sigma$ correspond to small sizes and high relative
momenta, so that
perturbative QCD can be used to calculate $P(\sigma=0)$.
The QCD result
\cite{BBFS93} for this in reasonable agreement with the value
extracted from data. This  continuity of $P(\sigma)$
indicates that the
transition between the
nonperturbative and perturbative regimes is
smooth. Furthermore, the probability is greater than 50\
for
a  nucleon and pion configuration to have a cross section greater than
average.

Color fluctuations of hadronic wave functions
account \cite{FMS93b} for
FNAL data for coherent neutron and pion
nuclear diffraction\cite{Ferbel1,Ferbel2}, as
discussed in Sect.~\ref{subs:ave}
Furthermore, the fluctuations of virtual photons
into hadronic
configurations of varying spatial sizes
have also been established experimentally via the
important role of PLC in yielding
approximate Bjorken scaling at
small x,  and the significance of large size configurations
in causing
the large diffractive cross sections
observed at HERA\cite{HERA}.

The role of color transparency in
nuclear quasielastic reactions is discussed in Sect.~\rep{sec:ct}.
Usually
initial and final state
interactions cause absorptive effects which reduce
the cross sections. If color transparency occurs,
such interactions are suppressed at high
enough Q$^2$. 
Examples with the potential for unambiguous
 interpretation include knocking out nucleons
\cite{Mue82,Bro82,FLFS88,jm90},
$\Delta$'s
\cite{FLFS89} or N$^*$'s
\cite{FS88,FLFS89,jm90,FS91,FMS92,FGMS92}.
CT requires the production of a PLC in two-body reactions.
The perturbative  and non-perturbative ideas about this are discussed.
However, even if a PLC is formed, it will expand
 as it moves through the nucleus\cite{FLFS89,jm90,jm91}. This expansion process
can be expressed through an effective PLC-nucleon cross section (imaginary
part of the scattering amplitude) $\sigma_{eff}$ which depends on the
distance $l$ the PLC has moved after the hard interaction.
In perturbative QCD one
can derive $\sigma_{eff}(l) \propto l/l_c$ for $l \ll l_c$ \cite{FS88}
and this behaviour seems to be a reasonable
     interpolation  for
 $l \le l_c$
as indicated by the observation of
precocious Bjorken scaling\cite{FS91}.
One can also express
expansion effects
in hadronic bases \cite{jm90,FMS92,jm91,jm92,FGMS92} with the result that
 $\sigma_{eff}(l)
\propto (l/l_c)^n$ with n=1-2 depending on the number of hadronic states
 used and assumptions regarding
the amplitudes of the soft interactions.
Similar expansion
effects occur in
non-equilibrium phase transitions.
(That the size of domain structures grows with time
as $\sqrt t$ as in quantum diffusion
 has been
found recently in the
Nambu-Jona-Lasinio
\cite {Be93}, and scalar $\phi^4$  \cite{Bo92} models.)
The expansion effect diminishes at very high energies when $l_c$ is large.
Thus initial and final state interactions are completely eliminated only at
high energies. At lower energies expansion and consequently absorption
takes place. Thus one unambiguous signature of color transparency
 is the predicted rise in the ratio
of the cross section to the plane wave Born approximation value
as the energy or Q$^2$ is increased.
The search for this distinctive property  may
be an effective
probe of the transition from nonperturbative to perturbative QCD.
The possiblilty of forming PLC at low values of
Q$^2\ge 1-2 GeV^2$ allows one to contemplate
searches for color transparency at
CEBAF. One must use double scattering
reactions to circumvent the substantial expansion
effects that are present. This is discussed in
Sects.~\ref{sec:ct} and \ref{sec:summary}

Color fluctuations can yield significant new
effects in heavy ion collsions, as described in Sect.\ref{sec:heavyion}.
The detection
apparatus can be set to trigger on the effects of
either very large or very small sized
configurations of the
initial state in the parton wave function of
colliding nuclei. Thus it is now practical to try to
observe
color transparency \cite{FS91a} and color opacity phenomena \cite{HBB91}.
Furthermore, using triggers for central nucleus-nucleus collisions may
be an effective
way to regulate the parton structure of colliding nuclei\cite{BF93}.
The function $P(\sigma)$, and its broadening as a
function of energy, is
important for understanding fluctuations observed in the recent CERN
heavy ion experiments
and for making predictions for RHIC and
LHC energies.  This subject is still in its
infancy.

So far only a very limited number of reactions sensitive to
color fluctuations has been studied experimentally.
At the same time, the  opportunities for progress are
enormous.
Sect.~\ref{sec:summary} describes a
number of reactions feasible for
CEBAF, BNL, FNAL, and HERA which will provide new insights into
strong interaction dynamics and hadronic structure.

\section{Coherence Phenomena in QED}
\label{sec:qed}
The dramatic consequences of
the increase of the coherence length with energy
may be surprising to some. But QED provides some relevant examples of
charge transparency, filtering and opacity.

\subsection{Charge transparency}

The effect of ``charge" transparency is well
known in QED where it was first suggested to weaken the ionizing
radiation of $e^+-e^-$ pairs moving through nuclear emulsion.
The pair could be produced by the
conversion of an ultra high energy cosmic ray photon
in the nuclear (A) Coulomb
field:  $\gamma +  A \rightarrow e^{+}e^{-} + A$ (Chudakov
\cite{Chud55}) and by
 decays of ultrafast pions $\pi^0 \rightarrow e^+e^-\gamma$
(Perkins\cite{Perc55}).
When the pair is close its point of production, the
pair interacts as a
very small dipole of
transverse size
$\sim R_A$
or $ 1/m_{\pi}$.
Such separations are much less than a typical atomic radius, so
there are very few ionizing interactions.
The
initial small  pair
separates as it moves and at larger separations, ionization does occur.
The ionization I has the approximate  form\cite{Chud55,ziel}
\bea
I=2 I_0 {ln (\theta x/r_{min})\over ln (r_{max}/r_{min})}
\label{eq:chud}
\eea
for small distances $x$ from the production point. Here $I_0$ is the
ionization produced by a single particle, $r_{min}$ is about an atomic
radius,
$\theta$ is
the initial angle between the positron and electron, and
$r_{max}$ is a screening length
($\sim 5 \times 10^{-7}cm$ for emulsions). More detailed calculations are
presented in Ref.\cite{moreqed}
Perkins' data was consistent with eq.\rep{eq:chud}
but, as reviewed by Zielenski\cite{ziel}, involved only seven events. Further
cosmic ray measurements confirmed this suppression
of ionization.
This charge
transparency effect has its hadronic analog in
deep inelastic scattering at small values of x, see
Sect.~\rep{subs:bj}.

\subsection{Charge filtering}
Another striking effect,
suggested by Nemenov \cite{Nemenov} and by Lyuboshitz and Podgoretskii
\cite{LyuPod},
concerns the survival probability of
positronium in its motion through a foil of density $\rho$ and length L.
The textbook definition of the total cross section $\sigma_{tot}$ for
positronium-atom scattering
involves an
exponential decrease
\begin{equation}
P = exp(-\sigma_{tot} \rho L). \label{expabs}
\end{equation}
But at sufficiently large
energies,
the coherence length $l_c$ of Eq.\rep{eq:lc} can
exceed  $ L$. Here
\bea
l_c\sim {2p_{lab} a_0^2 \over \sigma_{tot} \rho L} \label{posdis}
\eea
where a$_0$ is the Bohr radius of
positronium. The factor in the denominator enters because  the size of the
configuration decreases as it moves through the foil. Whenever
$l_c>L$
the probability for  positronium atoms to traverse a foil without
absorption decreases as a power of L, in strong contrast
with Eq.\rep{expabs}.

How does this come about? Consider the scattering of positronium by a first
atom, followed by propagation to a second atom and a second scattering.
At low and medium energies  it would be accurate
to treat the propagating $e^+e^-$ as the positronium ground state. This is
correct if the excitation energy is large
so that the appropriate coherence length is smaller than interatomic distance.
But at higher energies the
excited states propagate easily. Indeed one may
use completeness to sum over the excited states.
Thus if the first interaction produces a component
in which the e$^+$ and $e^-$ have a transverse (to
the positronium momentum)
distance separation
b, the pair is still at that same
separation when it encounters the second atom.

As a result the
amplitudes of diffractive high energy processes are expressed
as the
scattering amplitudes
of electron-positron pairs of transverse
separation b
averaged with the
product of the
initial positronium and detected final
state wave functions.
Thus,  neglecting  correlations
between atoms in the foil and the energy dependence of
$\sigma_{tot}$, one
finds that the probability for a transition to a
final state $f$ is given by:
\begin{equation}
P_{if} = \mid\int  \psi^{*}_{f} (r) exp [-\sigma_{tot} (b^{2})
\rho L/2] \psi_{pos} (r)  d^{3}r\mid^{2}, \label{pif}
\end{equation}
where
\begin{equation}
\sigma_{tot} (b^{2}) ={b^2 \over <b^2>} \sigma_{tot}
. \label{eq:sigpos}
\end{equation}
\noindent
How can we understand Eq.\rep{eq:sigpos}?
At high energies,
the cross section for positronium interaction with electrically neutral atom
is dominated by
two photon exchange diagrams. Furthermore, each interaction
with a target atom has a small momentum transfer.
The neutral system has no electric monopole term,
so each interaction is a
dipole, giving  a factor b.
A factor $\ln{{b^2\over <b^2>}}$ is ignored in Eq.
\rep{eq:sigpos}
for simplicity. In more technical language, the specific result
Eq.\rep{eq:sigpos} is the consequence of gauge invariance and
factorization.
The latter property is due to the rapid fall
off of the target atomic wave functions,
which allows an integration over the internal momentum
of the exchanged photons.

Suppose a positronium ground state is detected.
Then
Eq. \rep{pif} leads to the result
 \begin{equation}
P(positronium
\rightarrow positronium)=
16/(\sigma_{tot} \rho L)^{2}  \label{ptop}
\end{equation}
 for L such that
$2/(\sigma_{tot} \rho L) \ll 1$
and high energies satisfying Eq.\rep{eq:lc}\cite{FS91}.

Eqs. \rep{pif},\rep{ptop} are
different from ones of Lyuboshitz and Podgoretsky \cite{LyuPod}
  who suggested formulae
 for probabilities, but not for amplitudes as in
\rep{pif}. Eq. \rep{ptop} is remarkably different from
eq.\rep{expabs}  but it is applicable for
ultrarelativistic positronium beams only (cf. eq.
 \rep{posdis}).  Such beams have
been obtained recently from $\pi^0$ -- meson decays \cite{Afanas}.

The origin of this power-like fall off is that
positronium wave functions have significant probability to have b=0.
This reflects the
singular nature of the Coulomb potential at
small distances. An e$^+$e$^-$ pair with 0
transverse separation moves freely through the
foil
without losing energy. At the same time, the   foil
absorbs or strips projectile configurations of  large
size.
                    Thus in this process, the foil {\it plays role of
a filter for small
size configurations}. We shall discuss the QCD
analog of this in Sect.\rep{sec:soft}.
But other features of this positronium
example are also noteworthy\cite{FS91}.
Consider
an inelastic diffractive process
in which a final e$^+ e^-$ pair
is produced. Summing over f in Eq.\rep{pif} yields
a probability that varies as 1/L so that
\beq
{P(positronium \rightarrow e^+e^-)
\over P(positronium \rightarrow positronium)}
=(\sigma_{tot} \rho L)/8 , \label{ratiopos}
\eeq
for
$2/(\sigma_{tot} \rho L) \ll 1$.
 This is an example of  strong
enhancement of inelastic diffraction as
compared to elastic diffraction, which is also an important
feature of hadronic reactions. Another effect is
the
increase
of the average transverse momenta of $e^+e^-$ produced
in the dissociation of positronium, which is
analogous to pion diffraction
into mini-jets discussed in Sect.~\ref{subs:mini}
\subsection{Charge opacity}
Let us now consider effect of coherence for the
production of
lepton pairs $\it l^+ l^-$ in the propagation of
ultrarelativistic
positronium through media.  As noted above, the coupling of the
electrically neutral
positronium with the
photon is proportional to $b$,
so the probability to emit two
lepton pairs in collisions with
two atoms is proportional to $b^4$ (see Fig.~2). This
 is not equal to the square of the
probability to emit one pair in a collision with one atom. Therefore
more pairs are  produced by configurations in which
the electron and positron have a large
transverse separation.
Thus detecting two lepton pairs
selects ``opaque" - large  transverse
size configurations in positronium.
A similar effect for nucleus-nucleus collisions will be
discussed in Sect. \ref{sec:heavyion}.

\section{Color transparency in perturbative QCD}
\label{sec:ctpQCD}
The above treatment of coherent processes in QED is based on
three cornerstones: increase with energy of the essential
longitudinal distances, gauge
invariance and the factorization theorem.
All of these are also characteristics of QCD.
Thus the  essentials of charge transparency, charge filtering, and
charge opacity should find their analogs in QCD as color coherent phenomena.
However, QCD is  a nonabelian theory, with the properties of asymptotic
freedom and presumed infrared slavery,  so the analogy between charge and
color screening not complete.
Indeed, because of
the non-trivial interplay of perturbative
and non-perturbative
effects, as well as the high quality experimental
techniques required,
 the understanding of color coherent phenomena
has required a period of more than 20 years.

\label{sec:pqcd}
\subsection{Coherence length in QCD}

As discussed
in the introduction, the coherence length  is large at high energies.
This influences  hadronic and lepton interactions
with nuclei.
The presence of an important length for which coherent effects occur
invalidates a number of methods and ideas
which are often considered as cornerstones of high energy physics.
\begin{itemize}
\item[(i)]
It is quite popular to assume  that
fast projectiles  interact
consecutively with particles  at the same impact parameter
in making calculations for heavy ion cascade processes.
But actually a fast  projectile interacts
simultaneously with all of the target nucleons
within a distance $l_c$.
\item[(ii)]
The eikonal models currently used in describing
many high energy processes in particle and nuclear physics
assume that the intermediate state contains only
one fast hadron of typical size  $\sim $1 Fm.
But actually a coherent wave packet of hadrons/partons
can propagate for
distances up to the large $l_c$.
\end{itemize}

Inelastic states can be produced
without disturbing the target. At high energies
the longitudinal momentum transfer required to
produce
an excited  state of mass M$_X$, by a
high momentum $p_{lab}$ beam  particle of mass
$M_P$ is $\approx
{(M_X^2-M_P^2) \over 2p_{lab}}$
which is essentially negligible for all relevant
masses $M_X$.
This allows one to
make a sum over  the
amplitudes for producing various hadron states, i.e. to apply
closure.
To be specific, consider
the large cross section of inelastic diffractive processes that occurs at
high energy. Since $-q^2$ is very small,
the diffractive production of inelastic states
is not suppressed much by form factors. As a result, in high energy
processes, a coherent hadronic/partonic wave packet
propagates through
a nucleus. A single hadron does not propagate.
This physics has been understood before QCD
and implemented in the Gribov Reggeon
Calculus \cite{Gribov67}.

Diffractive hadron production is a significant effect in experiments.
To quantify this statement we present the numbers for the ratio
of single diffraction ($p +p \rightarrow  X+p$)
to elastic cross section at t=0 (at small
transverse momentum). Integrating the production amplitude over $M_X$ leads
at t=0 to
\beq
{{d\sigma^{single ~diff}\over dt}\over {d\sigma^{el}\over dt}}(pp)
 =0.2-0.3,     \label{e:frozen6}
\eeq
which is surprisingly large.
The importance of diffractive processes was
emphasized by the recent
discovery at  HERA\cite{HERA} of deep inelastic ep
scattering events with a
large rapidity gap in the hadron final state.
Such events
occur in the region of small Bjorken x, and
account for at least 5\% of the events with Q$^2\ge
10 GeV^2$. The properties of the events are
consistent with the
production mechanism being diffractive
dissociation of the virtual photon.

The ratio shown in Eq.~\ref{e:frozen6}
varies between 0.2 and 0.3 with energy as discussed
in  Sect.~\ref{subs:info}.
The large ammount of inelastic diffraction
made the elastic eikonal approximation  obsolete.
In particular, Gribov \cite{Gribov69}
 computed the inelastic
shadowing correction to the total cross sections of hadron scattering off
a deuteron. This
correction has been observed long ago, see
the review  \cite{Albery} and
Sect. ~\rep{subs:screening}

\subsection{Bjorken scaling for deep inelastic processes}
\label{subs:bj}
At sufficiently small x in deep inelastic processes
virtual  photon is transformed into hadron component
well before target because the correlation length
given by  $l_c={1\over 2m_N x}\approx .1/x  fm$
can be larger than the nuclear radius. The
discovery of approximate scaling for such values
of x  contradicted strongly pre-QCD ideas on
hadronic structure. This need for a fundamental
change began the history of color transparency.

Let us describe the history.
The discovery of Bjorken scaling for deep inelastic processes
at small x raised the question of
how to  match  the physics of hadron diffractive
processes with the physics of hard processes. Gribov \cite{Gribov69}
proved a theorem
which related
DIS electron scattering of nuclear target
with the process of $ e^+ e^- \rightarrow hadrons$.
The consequence for DIS was that if all hadron configurations
produced
interact with the target with the same
cross section,
the total cross section of virtual photon scattering off
sufficiently heavy target would be constant or
increase
with Q  instead of
decreasing as $ {1}\over {Q^2}$, as is observed.
 Thus
the seemingly natural hypothesis
that soft hadronic interactions
are determined by a pion cloud around the hadron and
are therefore mainly  universal
contradicts with the observed Bjorken scaling, and with
the parton model.
The initially computed lack of decrease with Q
results from high values of
transverse quark (anti-quark) momentum in the
necessary loop integral.
Bjorken \cite{Bj71}
suggested a resolution to this puzzle: if a photon
fluctuates into
a $q\bar q$ pair, the
quarks with large transverse momenta $k_\perp$
weakly interact with a target.
He obtained scaling by cutting off
(as in the parton model)
 the loop
integral at a value of $k_{max} \approx 300 $
MeV/c which is typical for a quark in a normal
strong interaction process.
Theoretical analysis of Feynman diagrams of
the fusion model
in which virtual photon fluctuates into a quark
antiquark pair (see Fig.~3)  found that in perturbative QCD
Bjorken scaling is obtained
because a configuration of
small transverse size b has a small cross section:
$\sigma \sim   \alpha_{s}(b)  b^2.
$
The quantity
$b$ is the transverse
separation between the q and the $\bar q$. The
factor $b^2$  arises as in QED, Sect.~\rep{sec:qed}
, but now
a color dipole of size b is involved.
The absence of interactions for small values
of b corresponds to an absence
of interactions for quarks at high $k_\perp$,
so
Bjorken scaling is restored\cite {FS88}.
Thus the analysis of fusion diagrams supports the  original
Bjorken suggestion that the dominant interactions
occur for q and $\bar q$ moving in a parallel
direction ($k_\perp$ is small)
and the corresponding dominance of the production of two jets aligned
along the direction of the initial virtual
photon.
The distinctive predictions of this picture are: one of the
jets carries
practically the entire momentum of the photon,
when the mass of diffractively produced
hadron system is comparable with $Q^2$, the correlations between the electric
charges and flavors of the fastest
and slowest diffractively produced particles are expected
to be the same as for the $ e^+ e^- \rightarrow$  hadrons.
 But
some
modifications are necessary to account for
gluon radiation
\cite{FS88}.

The typical virtual photons
of low x deep inelastic
scattering have a transverse polarization
\cite{anytext}. But for
longitudinally polarized
photons, the  $q\bar q$ parton model contribution is only
a next to leading twist effect,
and Bjorken's original logic is inapplicable.
In this case, the  leading twist contribution arises
when a longitudinally polarized
photon transforms into
a $q\bar q g$ component, with one of these partons
having a  small transverse
momentum. For such configurations, the
spatial distribution of color
is similar to that of ordinary hadrons. Thus
  the  spectacular
prediction of QCD is that diffraction
in deep inelastic processes is a leading twist effect,
 with a comparable
fraction of the total cross section as for soft
hadronic  processes. The
important difference between Bjorken's suggestion and
QCD is that hadronic states diffractively produced by
longitudinally polarized photons will have three  jets,
two with large transverse momenta.
Recent HERA data confirms the significant probability
of diffraction in deep inelastic processes\cite{HERA}.

A
number of authors used $\sigma \sim   \alpha_{s}(b)  b^2$
to suggest
 that the impulse
approximation in which a $q\bar q$ pair interacts
with a single nucleon at a time would be applicable for
the interaction of such a small
configuration with
 a nuclear target. QCD calculations are those of
Refs.~\cite{Bro82,Mue82,Bertsch81,Fra81}, and the constituent quark model
was applied in Ref.~\cite{zkl81}.
The idea was that exchanges by additional
gluons between current and target fragmentation regions
are reduced by an extra power of $ Q^2$.
However this should be modified,
since at small x  a small configuration may produce
large size color field. This idea can be
elaborated. The use of gauge invariance, asymptotic freedom and QCD
evolution equations allows one
to express the
cross section of scattering of small
transverse size $q\bar q$ pair off a
target T in terms of the gluon distribution of a target, $G_T$
\cite{BBFS93,FMS93}:
\beq
\sigma(b^2)={2\pi^2 \over 3}\left(b^2 \alpha_s(Q^2)\bar x
  G_{T}(\bar x, Q^2)\right)_{\bar x=1/s{b^2},Q^2=1/b^2} ,
  \label{sigmaqqbar}
\eeq
where $s=2 M_N p_{lab}$.
The presence of large size color
fluctuations in reflected in Eq. \rep{sigmaqqbar}
in the shadowing of the gluon distribution
which is expected to be
important for $x < 0.03-0.05$ (see
Sect.3.3.3).
Note that at fixed x, nuclear shadowing decreases
rapidly as
$Q^2$ is increased, according to the
QCD evolution equation.
Note also that the relevance of small x is enhanced by increasing
the beam energy, which leads to a
rather significant increase of $xG_N(x,Q^2)$ and
$\sigma(b^2)$, to the increase of shadowing for a nuclear target.

\subsection{Hard coherent diffraction}
\label{subs:hcd}
 The aim of this subsection is to demonstrate that
the factorization
theorem of QCD leads to effects in a wide kinematical region
which resemble  CT phenomena
expected in nuclear quasi-elastic scattering,
 but with certain distinctions. We also
explain that nucleus may play the role of filter
for ``small configurations".

The color fields emitted by closely separated systems of quarks and gluons are
cancelled for color singlet systems.  This effect of color neutrality
(or ``color screening")  leads to the suppression of initial
 or final state interactions, recall Sect.~\rep{sec:qed}
, if a small sized system or point-like configuration PLC
is
formed in a hard process. So far, the existence of  point-like configurations
is experimentally confirmed
only for virtual  photons, and
a clear observation of color screening and color transparency
would be an important clue to some poorly
understood problems of dynamics of bound states in QCD \cite{FMS92,FS88}.

The ideal situation to observe such effects would occur when it is
certain that PLC is formed, and when the energies are high enough so that the
expansion of the PLC \cite{FLFS88,jm90,jm91}
 does not occur (the ``frozen" approximation is
valid). We will consider two examples
of such diffractive processes: electroproduction
of vector mesons in QCD and pion dissociation into two jets.

\subsubsection{ Electroproduction of vector mesons in QCD.}

Let us first consider the simplest deep inelastic diffractive process
which can be legitimately calculated in perturbative QCD:
\beq
\gamma^*_{Longitudinal} + p \rightarrow~ Vector~ meson~ +~ p. \label{vlong}
\eeq
The process illustrated in Fig.~4 takes
place, sequentially in time, as
follows.
\begin{description}
\item { (i)} The virtual photon breaks up into a quark-antiquark pair
with a lifetime $\tau_i$ given by
\beq
q_+ \tau_i^{-1} = Q^2 + {k_\perp^2+ m^2 \over z(1-z)}
\approx Q^2 .\label{tauinitial}
\eeq
Here $m$ is the current quark mass, and z and $k_\perp$ are the
quark longitudinal momentum fraction and transverse momentum.
This estimate is valid for the production of a longitudinally polarized
vector meson only. In the case of a transversely polarized
vector meson,
an end-point non-perturbative
contribution arises from the kinematical
region where $z$ is close to 0 or 1: $z,1-z \sim  m^2/ Q^2$.
\item{(ii)}
The quark-antiquark pair then scatters off the target proton - this
cross section is given by Eq. \rep{sigmaqqbar}.
\item{(iii)} The quark-antiquark pair
then lives a time $\tau_f$ determined by
\beq
q_+ \tau_f^{-1} = {k_\perp^2+m^2\over z(1-z)}\label{taufinal}
\eeq
before the final state vector meson is formed.  We note that
$\tau_f \geq \tau_i.$
\end{description}
 Thus, the amplitude ${\cal M}$ can be written as a
product of three factors:
(i) the wavefunction giving the amplitude for the virtual photon to break
into a quark-antiquark pair - given by QED;
(ii) the scattering amplitude of the
quark-antiquark pair on the target  given by Eq. \rep{sigmaqqbar}; and
(iii) the wavefunction giving the amplitude for the scattered
quark-antiquark pair of flavor $f$ to become a vector
meson, V.

The latter overlaping integral depends on
\beq
\phi^V(Q,z) = \int^{Q^2} {d^2k_\perp\over 16\pi^3} \
\psi^V(k_\perp,z)\label{phiv},
\eeq
which is the  distribution amplitude for longitudinally
polarized vector mesons.
The restriction $k_\perp^2 < Q^2$ understood for the
integration in \rep{phiv}. This corresponds to
transverse q $\bar q$ separations
of   $ \sim 1/Q \approx 0$.
 This has
obvious implications for production from nuclear
targets, see Sect.~\rep{subssub:hard}.

The non-perturbative wave function $\phi^V(Q,z)$
is constrained by data.
The integral $\int_0^1 dz \phi^V(Q,z)$ is related to the
 the decay width for $V \rightarrow e^+e^-$, and
some extra information on $\phi^V(Q,z)$
is available from QCD sum rule analyses \cite{CZ}.
(For a critical discussion of the pion wave function at x=1/2,
see Ref.~\cite{BF89}.)

The resulting differential cross section for the reaction \rep{vlong}
can be expressed in terms of the  leptonic width,
$\Gamma_V$, the target gluon density, and a parameter
\beq
\eta_V\equiv\half
{\int_0^1\,{ dz \over z(1-z)} \phi^V(z)
\over \int_0^1\, dz \,\phi^V(z)},\label{etavdef}
\eeq
as
\beq
{d\sigma^{L}_{\gamma^* N \rightarrow V N}
\over dt}\bigg\vert_{t=0} = {
3 \pi^3 \Gamma_V m_V
\alpha_s^2(Q) \eta_V^2\mid
(1+i{\pi\over 2} {d\over d lnx}) xG_T(x,Q)\mid^2
\over \alpha_{EM} Q^6 N_c^2}. \label{dsigdtt}
\eeq
This result is based on a leading logarithmic
calculation (in ln (x)), so the  magnitude may not be
reliable. However, the Q$^{-6}$ dependence,   the
dominance of longitudinal polarization for both
the virtual photon and the produced vector meson
and the proportionality to the square of $x G_T$
and therefore fast increase  when x decreases
are firm predictions.

   Note that Eq.~\rep{dsigdtt}
has a different Q$^2$ dependence than
the 1/Q$^4$ of the non-perturbative
pomeron picture of Donnachie and Landshoff
\cite{DL}. The approach of Ref.~\cite{Brod94}
is closely related to that of Ryskin\cite{Ryskin}
who computed $J/\psi$ electroproduction in leading-logarithm PQCD,
but using constituent quark wave functions.
If the nonrelativistic $J/\psi$ wave function of Ref.~\cite{Ryskin}
is used in the equations of Ref.~\cite{Brod94} one obtains
a factor four less cross section than in  \cite{Ryskin}.
A constituent quark model
was used also by Kopeliovich et al\cite{KNNZ93}
to calculate electroproduction
of vector mesons.
However, QCD hard processes
should be computed through current
quarks and gluons but not through constituent quarks.
The constituent quark model used by \cite{KNNZ93}
is in qualitative variance with  QCD because of its
prediction that nuclear
shadowing for electroproduction of vector mesons is higher
twist effect.

The cross section for vector meson production
by transverse photons is expected
to be suppressed by extra power of
$Q^2$ which is consistent with the current
data \cite{NMC,fang} for $\rho$ meson production which
indicates that
longitudinal polarization dominates at large $Q^2$.

The
Eq. \rep{dsigdtt} is evaluated using $\eta_{\rho}=3.3-3.6$
of the QCD dispersion
sum rule wave functions of
\cite{CZ} and current parametrizations of
the gluon distributions in nucleon. The prediction for $Q^2= 10 GeV^2$ is:
${d\sigma^{L}_{\gamma^* N \rightarrow \rho^0 N}
\over dt}\Big\vert_{t=0} \sim {
16-23{\rm ~nb} \over {\rm GeV}^2} ,$
which is close to the preliminary diffractive $\rho$
leptoproduction cross section
of $14-27~ nb/GeV^{-2}$  reported
by NMC \cite{NMC}.
  This agreement indicates  that pQCD is relevant for
describing the interaction of small-sized q $\bar q$ pairs
with nucleons.
The above discussion also demonstrates
the feasibility of measuring  light-cone vector meson wave functions,
minimal Fock components of light -cone wave
functions of excited states : $\rho '$  etc.
No theoretical calculations of such
wave functions  exist at present. At the same time, the  calculation of
the overlapping integral
shows an extreme sensitivity
to $\psi'$ relativistic effects, so the
applicability of non-relativistic wave functions
to charmonium electroproduction is
 suspicious.

\subsubsection{  Coherent diffractive production of minijets
 by mesons}
\label{subs:mini}
The previous subsection was concerned with
studying
 small transverse size configurations in hadrons  (e. g. vector mesons)
 by selecting specific projectile characteristics
(large $Q^2$ longitudinal
photon). But one may
select small size transverse configurations in hadronic projectiles
by choosing instead a special final state\cite{FMS93}.
The simplest such reaction
is  a coherent nuclear process in
which a high-energy pion diffractively dissociates
into a $q \bar q$ pair of high relative transverse momentum $\vec k_\perp$
but the nucleus  remains intact (Fig.~5).

  If the final
$q\bar q$ pair carries of all of
pion's energy, only the
$q\bar q$ component of the light-cone wave function is needed for
calculations. In this case, only the small sized pionic configurations make
it through the nucleus without absorption and the nucleus can be said to
act as a filter.  (The use of the nucleus as a color filter has been
discussed for a long time,for the recent discussions see e.g.
Ref.~\cite{filter,FS85,Feinberg}, and can work if high enough
momentum transfers are involved.)
Observing high $k_\perp$ jets
insures that only small $q\bar q$ separations are involved
\cite{FMS93}.
The jets we consider involve high (greater than about 1 GeV/c) but
not very high values of $k_\perp$
less than
about 3 or 5 GeV/c (for currently accessible energies),
so we use the term minijets.
The mass of the final two jet system is  given by
$M_J^2=k_\perp^2/x(1-x)$, where
$x$ is the fraction of the beam momentum carried by the final state
quark and the anti-quark has a fraction $1-x$.
The effects of any soft particles of transverse momentum
have a negligible influence.
For
large $k_\perp$, $M_J^2\gg m_\pi^2$,
so the mass and
$k_\perp$ of the final state quark
-antiquark pair is significantly higher than
the masses and
$k_\perp$ of
the average pionic
configurations. Thus color is  sufficiently localized
and the system is small.
Furthermore, it is improbable
as a result of color screening and asymptotic freedom
 that many partons
(of non-zero longitudinal momentum) be
very close together, so restricting the calculation to include
the minimal number of constituents seems justified.
For larger values of $M_J^2$ than those
$M_J^2=k_\perp^2/x(1-x)$,
discussed here
the $qg \bar q$ jets ignored here could be
important and invalidate  the predictions
of this section.

The small sizes   and the large beam momentum greatly
simplify any computations of the matrix elements.
First, this  guarantees that
the interaction between the quark and anti-quark
must be  hard, and that using
the lowest order $pQCD$ diagrams
to calculate the interaction with a target is justified.
Calculation \cite{FMS93} shows that the $q\bar q$ interaction with the target
leads to an amplitude given by the momentum space pion wave function
$\tilde {\psi}_\pi(x,\vec k_\perp)$ multiplied by a factor of
$k_\perp^{-2}$.
Expressing the amplitude in
terms of the transverse q$\bar q$ separation
($\vec b$) shows that the
dominant contributions correspond to
$b\propto 1/ k_\perp$ \cite{FMS93}.
 Furthermore,
as discussed below, contributions from large values of $b$ occur with
higher powers of $k_\perp^{-2}$, and are of higher twist.
This is an important advantage of the hard processes discussed here.
For example in  computing the
elastic pion form factor $F_\pi(Q^2)$, all values of b lead,
in principle, to contributions to the leading $1/Q^2$ term.

The large beam momentum insures that
the momentum transfer to the high momentum
$q\bar q$ system is transverse ($t\approx -q^2_t$), so the $q\bar
q$-nucleon interaction is essentially independent of the longitudinal
momentum of $q\bar q$ pair. Furthermore, the frozen approximation, in which
the transverse $q\bar q$ separation $b$ is a constant, should be valid.

The above ideas allow an estimate of the
invariant amplitude ${\cal M}(N)$ for a
nucleon target which
is given at large invariant energies $s$
by
\beq {\cal
M}(N) = \int d^2b\psi_\pi(x,\vec b)\;{f(b^2)\over 2} e^{i\vec
k_\perp
\cdot\vec b}.  \label{e:M(N)}
\eeq
Here
$f(b^2)$ is the forward $q\bar q$ scattering amplitude normalized
according to the optical
theorem $Im\;f(b^2) = s\;\sigma(b^2)$ and
$\sigma(b^2)$ is given by \rep{sigmaqqbar}, and
$t=0$. The small real part of $f(b^2)$ is neglected.
The factor $e^{-i\vec k_\perp\cdot\vec b}$ accounts for the final
hadronic plane wave function of
two  jets
with a high relative momentum, so
Eq.\rep{e:M(N)} is given in an
approximation in which final state interactions
are neglected. Such can be included in
perturbative QCD and present estimates show that the effects
do not influence the A or $k_\perp$
dependence, but may  reduce the computed magnitude of
by a factor of two or so\cite{jm94}.
The assumed
$b^2$ dependence of the interaction $\sigma(b^2)$ allows an evaluation of
Eq.~
(3.10)
 in terms of the
Fourier transform $\tilde\psi_{\pi}(x,\vec k_\perp)$ by
expressing $b^2$ as $b^2 = -\nabla^2_{k_\perp}$.  Then
${\cal M}(N) = i\;{1\over 2}\;s\;\sigma(-\nabla^2_{k_\perp})
\tilde\psi_\pi(x,
\vec k_\perp)\;.$

The results of the analysis \cite{FMS93} show
that ${\cal M}(N)\propto
1/k_{\perp}^4
  \phi_{\pi}(x)$  for large values of
$k_\perp$. Here $\phi_{\pi}(x)$ is
the $q \bar q$ Fock component of
the pion light-cone wave function integrated over transverse momenta.
 The related cross section for
the diffractive production of minijets is a small but measurable,
${d^3\sigma_N\over dxdM^2_Td^2P_{N_t}} \approx 2\times 10^{-3}\;{\rm
GeV}^{-6}$. Uncertainties in input information regarding the
non-perturbative wave function, and gluon distributions cause
a factor of five or so uncertainty in the predicted
cross section. This seems to be the present state of the art accuracy.
Another problem is that  existing analyses ignore the
production of jets by the
target
Coulomb field. This amplitude
is proportional to the electric charge of
the target
 and is
singular when momentum transferred
to target tends to 0.

\subsubsection{  Hard Coherent processes with nuclei.}
\label{subssub:hard}
The dominance of configurations of small size
is a key feature of PQCD predictions for forward
diffractive leptoproduction of longitudinally polarized
vector mesons
with $1/2m_N x \gg 2R_A$ and for
coherent diffraction of pions into two jets.
Thus, even in a nuclear target, color
screening implies that the coherent $q \bar
q$ system can only weakly interact. In
leading-logarithmic approximation and in the light-cone gauge
only two gluons
connect the photon-vector meson system
(pion - two jet system) with the nucleus, as illustrated in
Figs.~4 and 5.
Thus  the hadronic system propagating through the nucleus
suffers no initial-state or final-state absorption, and the
nuclear dependence of
the $\gamma^* A \rightarrow V A,
\pi +A
 \rightarrow $ 2 jets + A
forward amplitude will be approximately additive in the
nucleon number $A$.  We can also understand this remarkable feature of
QCD from the space-time arguments given above: the final state vector
meson is formed over a long time
$\tau_f,$
from a compact $q \bar q$ pair
which does not attain its
final physical size and normal
strong interactions until it is well outside the target
nucleus.

Although the vector meson (two jets) suffers no final state interactions, the
forward amplitude is not strictly additive
in nuclear number since the gluon distribution itself is shadowed.
Thus one predicts an
 identical nuclear dependence for the forward vector meson
diffractive
leptoproduction cross sections,
diffractive production of dijets by pions, the
longitudinal structure
functions $F^L_A(x,Q),$ and the square of the gluon structure functions:
\beq
{{d\sigma\over dt}(\pi A \to 2 jets + A)\big\vert_{t=0}\over
{d\sigma\over dt}(\pi N \to 2 jets + N)\big\vert_{t=0}} =
{{d\sigma\over dt}(\gamma^* A \to V A)\big\vert_{t=0}\over
{d\sigma\over dt}(\gamma^* N \to V N)\big\vert_{t=0}} =
\left [{F^L_A(x,Q) \over F^L_N(x,Q)}\right ]^2
= {G^2_A(x,Q) \over G^2_N(x,Q)} = A^{2 \alpha_g(x,Q)}\ .\label{ratios}
\eeq
(Note that at finite energies, an
interpolation to the unphysical $t=0$ kinematical point is needed.)
The nuclear gluon distribution is expected to be more strongly shadowed
than the nuclear quark structure
functions at intermediate $Q^2$ because of the larger color
charge of the gluon. Numerical estimates
\cite{FS88,fsl} lead to
$xG_A(x, Q_0^2)/AxG_N(x,Q_0^2) \sim 0.7-0.8~~
(0.4-0.5)$ for $A=12~(200)$ and  $x \sim 0.01-0.03$,  a result
which seems to
be supported by the recent FNAL data of E-665 \cite{Melanson}.
However, at fixed $x \sim 0.01-0.03,$
shadowing substantially decreases with
$Q^2$ due to scaling violation effects \cite{fsl}, which should
lead to an effective increase of
transparency for $\rho $ leptoproduction
at fixed $x$ with increasing $Q$.
In the case of the $\pi +A \rightarrow 2 jets +A$
process the corrections to the main
cross section $\sim A^{1/3} \over p_t^2$ were
found \cite{FMS93} to be positive leading
to a noticeable increase of the cross section at intermediate
$k_\perp$.

 The  calculations described above are applicable
to the  near forward production of
vector mesons or  2 jets. Obviously the A dependence should change with t.
At  $-t R_A^2/ 3\ll 1,$ (within
the diffractive peak)  an additional A dependence is given by the square of
the nuclear electromagnetic form factor.
This steep decrease with t  can be used as experimental signature of
coherent processes.
However, for $-t R_A^2/ 3\gg 1$,
incoherent processes, in which the
leading hadronic system  is accompanied by the production of
other hadrons from nuclear disintegration,
will dominate the cross section.
The existence of nuclear  shadowing implies that
gluons at small $x$ cannot be associated with individual
nucleons.  Thus one can have events
where momenta $-t \ge 0.1~GeV^2 $ are  transferred  to
each of
several  nucleons
which subsequently fragment.
This is expected to lead to a slower $t$-dependence of the cross section and
a smaller energy transfer per interacting
target nucleon than for scattering from
a single nucleon.
The expected $A$-dependence is intermediate between
that expected for shadowing of $G_A(x,Q^2)$
and $A$.

The recent nuclear target $\rho$ leptoproduction measurements
from the E-665 experiment\cite{fang} appear to indicate onset of the color
transparency predicted by PQCD for
incoherent $\gamma^* A \to \rho N(A-1)'$  reactions.
The onset of this phenomena occurs again at  $Q^2 \simeq$ a
few GeV$^2,$ the same scale at which Bjorken scaling is observed in deep
inelastic lepton scattering reactions.
Preliminary
data
\cite{NMC} from the NMC also
confirm  higher
values of the transparency ratio for $Q^2 \ge 3~ {\rm GeV}^2,$
observed in \cite{fang}
although the NMC data do not indicate
a $Q^2$ variation of transparency in
their  $10 \ge Q^2 \ge 3 ~{\rm GeV}^2$ range.
Note that the
high-$Q^2$ NMC data correspond to
$x$  large enough for the
essential longitudinal distances to be smaller
than the nuclear diameter, so the
transparency here would be influenced by
the deeper nuclear penetrations
of virtual photons compared with hadrons.
Thus, one needs to be cautious in interpreting these
data directly in terms of PQCD color transparency of the
outgoing $\rho.$
As we have emphasized,
the nuclear dependence of forward diffractive $\rho$ leptoproduction
in which the target nucleus is left in its ground state
can provide a decisive test of
color transparency.

We conclude this section by discussing color screening phenomena in
nuclear charmonium photo-production at energies
$E_{\gamma} \le 20 GeV$ \cite{Farrar90}. Here
$l_c$ is smaller than the mean internucleon distance of about 2 fm.
For these kinematics, $\Psi,\Psi'$ are produced in configurations of small
size $r_t \sim 2/m_{\Psi} ((2/m_{\Psi'})$  leading to the prediction
$R=\sigma^{VDM}_{\Psi' N} /\sigma^{VDM}_{\Psi N}
 \sim \sim m^2_{\Psi}/m^2_{\Psi'} \approx 0.7 $
\cite{FS85}, in reasonable agreement with
experiment. Here $\sigma^{VDM}$ is the cross
section
 obtained from the vector dominance
model analysis of the photoproduction, for a review
of the VDM
see e.g. \cite{Feynman}.
The naive  expectation for this ratio
would be $R= (r^2_\Psi$')/$(r^2_\Psi) \approx 4 $.
Due to the smallness of $l_c$, the   $\Psi, \Psi'$  expand rapidly and are
 absorbed at these energies with the genuine $\Psi (\Psi') N $ cross
section. Indeed
Farrar et al \cite{Farrar90} find
from the analysis of SLAC data on $\Psi$ nuclear photoproduction
that $\sigma(\Psi N) \approx 3 mb$ as compared
to $\sigma^{VDM}(\Psi N) \approx 1 mb$
from application of the vector dominance
model for photoproduction of hydrogen. It would be extremely interesting
to study the A-dependence of $\Psi'$ photoproduction at these energies to check
the
color screening prediction that $\sigma(\Psi' N) \approx 10 mb$. It is also
important to develop an adequate theoretical approach for considering
nuclear photoproduction of heavy mesons for the transitional
range of energies $20 \le E_{\gamma} \le 200 GeV$
where $2 fm \le l_c \le R_A$.

\section{Soft diffractive physics and color fluctuations}
\label{sec:soft}

Consider inelastic diffractive processes in which an
incident beam is excited, but the target is not.
If such a process is to occur one must
introduce
components in the projectile interacting with different cross section.
 To illustrate this point,
consider a two component model of the projectile a hadron h
with $\mid h> =a\mid \alpha> +
b\mid \beta>$.
After passing through an absorber the wave packet
 is modified
to    $\mid h> =\epsilon_1 a\mid \alpha> + \epsilon_2
b\mid \beta>$. If two components are absorbed with equal strength,
$\epsilon_1=\epsilon_2$
     and the final state is just $ \epsilon_1\mid h>$ and no
inelastic states are produced. On the other hand if $\epsilon_1 \neq
\epsilon_2$ the final state does not coincide with $\mid h>$ and inelastic
diffraction takes place \cite{FP56,GW60}.
The consequence of large inelastic diffraction cross
sections is that hadrons must have significant fluctuations
between configurations of different interaction strengths (and
via results like Eq.~\rep{eq:sigpos}) sizes.
What are dynamical mechanisms of fluctuations of the cross sections?
One is color neutrality which
we discussed in the previous sections - for
small size fluctuations interaction is smaller  because
fields of individual
closely separated quarks and gluons
cancel each other.  Another  mechanism is
that nucleon configurations with and without
a pion cloud have different interactions with the target.
Since color dynamics determines all the interactions, we use the term
{\it color fluctuations.}

Understanding such fluctuation effects is simplest at high
energies, for which the coherence length of Eq.~\rep{eq:lc} can be
larger than the nuclear diameter. In that case
one may treat h as frozen in its initial configuration \cite{Mandelstam}.

\subsection{Scattering eigenstates formalism}

It is tempting to model the wave functions of the hadronic
fluctuations and the interaction cross sections.
For recent  attempts
using the two-gluon exchange interaction with Gaussian
constituent quark model wave functions see Ref.~\cite{bk} and
references therein.

But in general, many degrees
of freedom are involved and model dependence is unavoidable.
Instead, the physics of color fluctuations  can be
incorporated
by  applying the Good and Walker
formalism of expanding the beam wave function as a sum of eigenstates
of the purely imaginary scattering amplitude T.
Thus the hadronic state $\ket{\Psi}$ is given by
 $ \ket{\Psi} = \sum_k c_k \ket{\psi_k}$ where
  $  {\rm Im} T \ket{\psi_k} = t_k \ket{\psi_k}$, and the
normalization is
$\sum_k |c_k|^2 = 1$.
Thus of the
optical theorem ($ {\rm Im} T|_{t=0} = \sigma$) allows a
high energy treatment of the projectile
as a coherent superposition of
scattering eigenstates, each with an eigenvalue $\sigma$. The probability that
a given configuration interacts with a nucleon with a total cross section
$\sigma$ is $\ps=\sum_k |c_k|^2 \delta(\sigma-t_k)$.
 For many high-energy applications
it is sufficient to know $\ps$ and not the more complicated $t_k$.
In particular,
the small t elastic and diffractive scattering cross sections are given by
\bea
  \frac{d\sigma_{el}}{dt} =\frac{1}{16\pi} \left(\sum_k |c_k|^2 t_k\right)^2
                  = \frac{1}{16\pi} \av{{\rm Im} T}^2=
\frac{1}{16\pi}
 \av {\sigma^2}
,  \label{eq:sigel}
\eea
and by applying closure over diffractively produced hadron
states
\bea
\left(\frac{d\sigma_{diff}}{dt}\right)
= \frac{1}{16\pi}
 \left( \av{{\rm Im} T^2}- \av{{\rm Im} T}^2 \right)
                            = \frac{1}{16\pi}
                              \left( \langle \sigma^2 \rangle
                                    -\langle \sigma \rangle^2
                                     \right).         \label{e:diffmp}
\eea
The last result is the
Miettinen and Pumplin\cite{MP78} relation, which
 reaffirms the close connection between color fluctuations
and inelastic diffraction.

We stress that Eqs. \rep{eq:sigel} and \rep{e:diffmp} are
valid only for forward scattering where $-t$ is very small
$(\ll R_N^2$).
Otherwise, one must include the t-dependence of $t_k$ which
 varies  with $k$. In that case, the
coherence between contributions  of states with different masses is
lost.

\subsection{Current information on color fluctuations in hadrons}
\label{subs:info}

What is known about $P(\sigma)$?
It is convenient to consider moments:
$\langle \sigma^n\rangle =\int d\sigma \s^n P(\sigma)$.
The zeroth moment is unity, by
 conservation of probability, and the first is the total cross section.
The second moment has been  determined using current diffractive
 dissociation data
from the  nucleon using Eq. \rep{e:diffmp}.
Complementary determination is possible
 using information on  inelastic corrections to the total hadron-deuteron
cross section. These two determinations give consistent values of
 $\langle \sigma^2\rangle$ corresponding to the variance of the distribution:
\beq
\os(p) \equiv \left( \langle \sigma^2 \rangle
                                    -\langle \sigma \rangle^2
                                     \right) /  \langle \sigma^2\rangle \sim
0.25
, \os(\pi) \sim 0.4,
\eeq
 for $ p_h \simeq$  200~ GeV/c.
Current diffractive data
indicate that $\os(p)$ increases
with
$E_{inc}$
up to $\sqrt{s} \sim 100 GeV$ reaching
values around 0.35, and probably starting to drop at
higher energies  approaching $\sim 0.2$ at $\sqrt{s} \sim $ 2 TeV,
 see Ref.~\cite{BBFS93} for a review and references.

The value of
 $\langle \sigma^3\rangle$
has been
determined from
the analysis of deuteron (D) diffraction dissociation
data:
$p + D \rightarrow X + D$.
The result
corresponds
 to $\langle (\sigma -\langle\sigma\rangle\rangle^3 \approx 0$
as would occur for a distribution nearly symmetric
around $\langle \sigma\rangle$\cite{BBFHS93}.

For small values of
$\sigma$
further information
can be obtained from perturbative QCD, which implies \cite{FS91a}
\begin{eqnarray}
P(\sigma) \sim \sigma^{N_q - 2}, \label{e:psigsmall}
\eea
for  $\sigma \ll <\sigma>$,
where $N_q$ is the number of valence quarks. Thus the nucleon
distribution $P_N(\sigma)$
is $\sim \sigma$ for small
$\sigma$ while for the pion $P_{\pi}(\sigma) \sim $ constant.
The results of reconstructing
$P_N(\sigma) $ and $P_{\pi}(\sigma)$ from the first few moments of
$P(\sigma)$ and eq \rep{e:psigsmall} are shown in Fig.~6.
They indicate  a broad distribution
for proton projectiles and
an even broader one for pion projectiles. One
also expects even further broadening for
K-meson projectiles.

\subsection{Small size configurations in
the pion}

One may test this approach by using
QCD
to compute
$P_\pi(\sigma=0)$
at high energies.
 Indeed
the  physics at small $\sigma$ is
dominated by small-size valence $q \bar{q}$ ~Fock state configurations of the
pion, measured by the wave function
$\Psi(x, b)$,
 where $x$ is the
light-cone momentum fraction, and
$b$
is the transverse
 distance between
quark and antiquark.
Thus
\begin{eqnarray}
   P_\pi(\sigma=0) = \int_0^1 dx |\Psi(x,b = 0)|^2
\frac{1}{4}
        \left(\frac{ d b^2 }{d\sigma}\right)_{b=0}.
            \label{e:psig0}
\end{eqnarray}
The asymptotic QCD wave function of the pion
is normalized to reproduce the muon decay width of the pion
and is given by Ref.~\cite{BL80} as:
\begin{eqnarray}
  \Psi(x,b = 0) = \sqrt{48}\pi x (1-x) f_\pi,
                                  \label{e:psirt0}
\end{eqnarray}
where $f_\pi$ = 93 MeV is the pion-decay constant.
Then
\begin{eqnarray}
P_\pi(\sigma=0) = \frac{2\pi^2}{5}f_\pi^2
\left(\frac{db^2}{d\sigma}\right)_{b =0}.
\label{e:est}
\end{eqnarray}
If the
more realistic Chernyak-Zhitnitski wave function \cite{CZ}, based on QCD sum
rules, is used, the value of $P_\pi(\sigma=0)$ is increased by
a factor of 25/21.

Since color effects dominate for small transverse separation, $b$,
of the valence $q \bar q$ pair in the pion, we can in fact apply perturbative
QCD directly to calculate $\sigma(b)$ at $b^2 \ll \langle
b^2\rangle$ . For small-size configurations, the dominant process in
$\pi$N scattering is exchange of two gluons, a process analogous to the box
diagram  Fig.~3,
 for $\gamma$-hadron scattering, where instead of the $\gamma \to q
\bar q$ coupling, the $\pi\to q \bar q$ wave function enters.  Taking the
leading gluon exchange diagrams into account in the limit in which the center
of mass energy $\sqrt s$ is large compared with $b^{-1}$, and using
Eq.~\rep{sigmaqqbar}, we find
\cite{BBFS93},
\begin{eqnarray}
P_\pi(\sigma=0)=\frac{3f^2_{\pi}}{5 \alpha_s(4k^{2}_{t})\bar x
  G_{N}(\bar x, 4k^{2}_{\perp})},
\label{sigmapi}
\end{eqnarray}
where $\alpha_s(4k^{2}_{t})$ is the QCD running coupling constant, and
$G_N(\bar x,4k^{2}_{t})$ is the gluon distribution in the nucleon;
 $\bar x= 4k^2_\perp/s_{\pi N}; k^2_\perp \approx 1/b^2$ where $f_{\pi}$ is
 the constant for $\pi \rightarrow
\mu \nu$ decay. For realistic
value of $\alpha_S(1 GeV^2) \approx 0.4 $ we find
$ P(0) \sim 0.017-0.023 mb^{-1}$
which is quite close to the value of about 0.015-0.02      obtained
 from fitting $\ps$.
Since $G_N(x, Q^2)$ can
be measured independently, Eq. \rep{sigmapi}  is  exact
(in the leading logarithmic approximation), parameter-free
result in the limit $1/s \ll b^2\ll \langle b^2\rangle$.  Since the
gluon distribution substantially
increases with decrease of x at fixed $Q^2$ Eq.
\rep{sigmapi} indicates that  probability to find a hadron in configuration
with $\s \le \s_0 \ll  \langle \sigma^2 \rangle$ decreases with increase of
energy.

\subsection{Nuclear inelastic diffraction as evidence for color fluctuations
 near the average value}
\label{subs:ave}
Nuclear inelastic coherent diffractive hadron
production provides another
nontrivial experimental test of the idea of color fluctuations.
In this case, the target nucleus remains intact
as the beam diffractively dissociates. The total
diffractive cross section can be computed
using $\ps$  \cite{FMS93b}:
\bea
&\sigma_{diff}(A)=\int d^2B &\nonumber\\
&\left[ \int d\sigma
P(\sigma) \sum_n  \left[<h\mid F(\sigma,B) \mid n>^2 \right]-
\left[\int d\sigma
P(\sigma) <h \mid F(\sigma,B) \mid h>
\right]^2\right],&\qquad \label{cohdif}
\eea
 where
$ F(\sigma,B) = 1 - e^{-{1\over 2}\sigma T(B)}$
and $T(B)=\int_{-\infty}^\infty \rho_A(B,Z)dz$.
Here the direction of the beam is $\hat Z$ and the distance between the
projectile and the nuclear center  is $\vec R=\vec B + Z\hat Z$.
The advantage of Eq. \rep{cohdif} as compared to  the related equation of
Ref.~\cite{zkl81}   within the
constituent quark two gluon exchange model
and some related works reviewed in Ref.~\cite{shocked}
is that there is no need to assume the validity
of pQCD at average interquark distances
in hadrons,
 to introduce
gauge non-invariant effects due to using
 a  nonzero gluon mass or, to
 use  constituent quark model wave
functions.

It is instructive to consider the extreme
 black disk  (bd) limit of
this formula (which would be a reasonable model if
color fluctuations were small). In this limit,
the function $F(\sigma,B)$ is unity
  for positions inside the nucleus and  zero otherwise,
so that $\sigma_{diff}$ vanishes!  In particular,
the black disk model leads to the
expressions
$\s^{bd}_{tot}=2 \pi R_A^2$, $\s^{bd}_{el}=\pi R_A^2$ and $\s^{bd}_{diff}=0$.
Another way to show that color fluctuations cause
$\sigma_{diff} (A)$
is to ignore the composite nature of the nucleon and take
$\ps$ to be  a delta function,  e.g.
 $\ps =\delta (\sigma - \s)$ gives
a zero result  for $\sigma_{diff}(A)$.
This is in marked contrast with the calculation
 using Eq. \rep{cohdif} and realistic $\ps$ which leads to
 $\sigma^{pA}_{diff}(A) \propto
A^{0.80}$ for  $A \sim 16$ and
$A^{0.4}$  for $A \sim 200$.
For the pion projectile we find even faster A-dependence.
 The results of the calculation agree well with the A-dependence
 of semi-inclusive data  of \cite {Ferbel2}on $n+ A
\rightarrow p\pi^- + A$ for $p_n \sim 300 GeV$ and of
\cite{Ferbel1} on  $\pi^+  + A \rightarrow \pi^+  +\pi^+  +\pi^-  +A$
 for $p_{\pi^+}$ = 200 GeV, see
Fig.~7.
In both cases fluctuations near
the average
 value of $\sigma$  dominate. The different $P(\sigma)$ of Fig.~6 lead to
the narrow bands shown in Fig.~7.
The present  numerical results (using $\ps$) for $\sd(A)$
are reasonably close to those
of the preQCD model of \cite {BS83}. This is
because similar  values of $\os$ were used.

In the limit that A becomes infinite,
the nucleus acts as a black disk except for  configurations of very
small size.
 Indeed Ref.~\cite{Bertsch81,zkl81}
suggested that the effects of such small-sized
configurations would dominate the  pion-nucleus
 inelastic diffractive cross section.
This early result is inherent in Eq. \rep{cohdif}.
 The integration over small values of $\sigma$ in
Eq.\rep{cohdif}  gives a result $\propto 1/T(B)$ for
 values of
 B within the nucleus.
 The integration over $d^2B$ leads to $\sd\propto R_A \propto A^{1/3}$.
Numerical evaluations of Eq.\rep{cohdif}  show that
the behavior is close to A$^{0.33}$ for fantastic  values of A
greater than about 10000.

An additional
 indication  that small-sized configurations are
not important here is that
the contribution to $\sd$ for values of $\s < 5 $
mb varies from $2\%$ to $ 5\%$ as A increases from about 12 to 200.
Thus the pQCD contribution to inelastic diffraction suggested   in
 Ref  \cite{Bertsch81,zkl81} doesn't
exceed $2\%$ to $5\%$.

\subsection{Inelastic screening corrections to the
total cross section}
\label{subs:screening}

The total nucleon-nucleus cross-sections
provide another set of
observables that are sensitive to the effects of
the presence of color
fluctuations.  In Glauber theory an incident
hadron remains in that state after interacting
with a nucleon. But actually a coherent superposition of states propagates
 and interacts. This is the
inelastic
screening (shadowing) process \cite{Gribov69} in which
the incident nucleon diffractively dissociates
into an excited or
resonant state, N$^*$, by its
interaction with a nucleon in the target and de-excited by another.
This correction is closely related  to color
fluctuations. The absence of color fluctuations
correspond to an absence of diffractive
dissociation, recall Eq. \rep{e:diffmp}, and an
absence of inelastic shadowing. Furthermore,
forward diffractive dissociation
cross-sections are used as input in computations of
inelastic shadowing effects, Ref.~\cite{karmanov}.

The presence of
inelastic shadowing has been established for a
long time \cite{murthy,Albery}. However, this effect is
determined by subtracting the computed
standard multiple scattering contribution from the
measured
total cross sections.
Much of the input data to the standard calculation
has changed with time, so Jennings and Miller
\cite{jm94b} used
updated
versions of the nuclear densities, nucleon-
nucleon forward scattering amplitude, and diffraction dissociation cross
sections. The newer calculations
reaffirmed the need to include inelastic shadowing
corrections of roughly the size found in Ref.~\cite{murthy}.
The results of \cite{jm94b} were in contrast with a claim
by
Nikolaev\cite{nikolaev},
that the total cross section data can not be reproduced without
allowing the nucleon-nucleon total cross section to be significantly
(5-15\% ) larger for bound target nucleons than for free target nucleons.
This conclusion is based on
the assumption that
any small differences between
the calculation
of nucleon multiple-scattering plus inelastic shadowing
are due only to enhanced nucleon-nucleon cross sections.
But the total neutron-nucleus cross section is not very
sensitive to the total nucleon-nucleon
cross section, so that relatively large
medium effects are required.
 But
the small differences between
calculations and the data could be due to any number of other
effects\cite{jm94b}.

To summarize Sect.~4,
the presence of significant color fluctuations in hadrons
 is well established now. Furthermore,
the  Reggeon Calculus indicates, via the AGK
cutting rules \cite{AGK},
 that diffractive
and high multiplicity  processes
 correspond to different cuts of the same diagrams. Consequently,
the observation
of significant inelastic coherent diffraction as well as of inelastic
 shadowing correction to $\sigma_{tot}(hA)$ implies that color fluctuations
 should also play  significant role in
inelastic
 hadron collisions.

\section{Color Transparency in nuclear quasielastic reactions}
\label{sec:ct}

If color transparency CT occurs, initial and/or final state
interactions in large Q$^2$ nuclear
 quasielastic reactions vanish or are suppressed.
This color transparency requires the formation of a small sized object
in  two-body projectile-nucleon collisions. Thus
the experimental resolution
must be good enough to insure that no extra pions are produced
and the energy transfer to  the recoil nuclear system is small ($\le 70-100
MeV$).
This requirement, stringent at high  energies has hindered progress in this
field.
The CT idea is based on: small short-lived color singlet objects are
produced in elastic hadronic reactions at high momentum transfer Q$^2$;
such objects have small interactions with nucleons;   the small system is
not an eigenstate so, unless its energy is  very high,
it expands and interacts  as it moves through the nucleus.

\subsection{Is a small system made?}
\label{subs:small}

Perhaps the most interesting question is whether or not a small
system is made in a high Q$^2$ hadronic exclusive process.
The present postulate is
that
at high Q$^2$, the
matrix elements are supposed to be dominated by components or
configurations that behave as of  smaller than average
size.
Such small-sized configurations or wave packets
have been termed
point- like
configurations PLC and we use that term as a convenient
abbreviation.

If one considers asymptotically large values of Q$^2$,
perturbative QCD
holds and
PLC are produced\cite{BL80,Mue82,muellerpr}. The
argument is that e.g. elastic electron-pion
scattering proceeds by the absorption of a virtual photon by a
quark.
This  struck quark is off the energy shell by $\Delta E \sim |\vec q|=Q$,
 so it has a lifetime
$\tau$
determined  by the uncertainty principle $\tau=\hbar /Q.$
The virtual quark must decay by emitting a particle (gluon), but
the  final state can be a pion only if
the other quark is close enough (within
$r\simeq c \tau$ ) to absorb the gluon.
Thus the system's size
is of the order of $\hbar$ ~c/Q.

More formally, the lifetime
can be computed in terms of
$x_i$,
the longitudinal momentum fraction,  and  the
transverse momentum of the
interacting
quark before and after the photon is absorbed.
The relevant hard scattering operator T$_H$ can be expressed as a
function of the
transverse separation $\vec{b}$ between the quarks by taking the Fourier
transform. The result is that $T_H\sim \alpha_s ((1-x_i) Q^2)
K_0(\sqrt {1-x_i} Qb) \sim
\alpha_s e^{-\sqrt {1-x_i} Qb}$.
This means the
quarks must be within a transverse separation of $1/\sqrt{1-x_i}Q$.

Eloquent criticisms
of  early pQCD  calculations
were put forward by Radyushkin\cite{ar}  and Isgur and Lewellyn Smith
\cite{isgurls}. Those authors
argued
that the dominant contributions to the integrals over $x_i$
occur for
small values of
$1-x_i$.
 In that case $\alpha_s$ is to be
evaluated at very small arguments where it is much too large for
perturbative calculations to have merit.
This
physics is closely related to
the Feynman
mechanism \cite{Feynman} in which  essentially the entire
longitudinal momentum of initial
hadron
is carried by a single quark that is turned upon absorbing  the photon.
Then, small values of
$(1-x_i)$
may enter in the
argument of the $K_0$ function
and  small transverse separations are not relevant.
Moreover calculations using the non-perturbative methods
can reproduce form factors, see e.g.
Bakulev and Radushkin \cite{br} who find a mechanism similar to that of
 Feynman
to dominate the pion form factor for values of $Q^2$ up to 10 GeV$^2$.
Braun and Halperin \cite{BH} used a
refined dispersion sum rule
approach to show
that such a mechanism dominates
up to Q$^2$ = 10 GeV$^2$, but also that the contributions come from
small transverse separations of the order of 0.2 Fm.

But the pQCD arguments and the criticisms thereof were
not complete because one must
investigate the possible role of
low momentum (soft) long wavelength
gluons that are radiated as the colored quarks are
accelerated. This is analogous to the bremsstrahlung that accompanies the
acceleration of an electron, so the
effects of such radiation can be included
via the a form factor similar to that introduced by Sudakov
\cite{sudakov},
which
 decreases the probability for elastic scattering of a free fermion.
However,
for a color singlet system, recall Sects.\rep{sec:intro},\rep{sec:qed},
\rep{subs:hcd},
 the gluon radiation contributions
cancel if the quarks and gluons making up the system are closely separated.
The net result is that significant
contributions to the elastic form factor
tend to occur for configurations of small size.

The so-called Sudakov effects were known early on
\cite{muellerpr,BL80},
but
numerical evaluations
  did not occur
until recently with the work of Botts, Li and
 Sterman\cite{listerman,Li,Botts,sterman}.
The key technical point was to incorporate
Sudakov effects
via a pion wave function
depending  on b,
which emphasizes small values of b,  and
one evaluates
$\asmu$ at the larger
of $x_1x_2 Q^2$ or
$1/b^2$. The net result is that
such pQCD calculations of the pion and nucleon form factors are
more
consistent
theoretically and also agree with experiment within a factor of two or so.
There are
also
higher order and higher  twist correction
terms, see the summary in Ref.~\cite{jp}.
The distribution of relevant transverse separations  b within the
pion and nucleon
wave functions obtained within this modern pQCD
approach is rather wide\cite{Li}.
Note however that the use of similar logic, as applied by
Sotiropoulos and Sterman \cite{SS}
to  elastic pp
scattering at large s and $-t \ge few ~GeV^2$,  leads to the conclusion that
the observed t-dependence of the data
could be reproduced only if the size of the $3q$ configuration is less than
0.4 of the average nucleon size. If so, Sudakov effects would be negligible
in the nucleon form factor up to $Q^2 \sim 10 GeV^2$.
See also Ref.~\cite{BK94}

Thus different approaches to understanding hadronic wave functions
yield very different
b-distributions. More definitive work is necessary.

So far we have discussed pQCD calculations. But if one is interested in
seeing how color transparency effects grow as Q$^2$ is increased from
low values it is necessary to see if
non-perturbative calculations also admit a PLC.
The current authors have examined\cite{FMS92,FMS931}
  several different models by
considering  a ratio defined as $b^2(Q^2)$:
\bea
b^2(Q^2)\equiv{\langle N(\vec q)|b^2  T_H|N(\vec 0)\rangle \over
\langle N(\vec q)| T_H|N(\vec 0)\rangle}.
\label{eq:bsquared}
\eea
The above notation specifies that a nucleon at rest absorbs a momentum
$\vec q$, and the denominator is  the nucleon form factor $F(Q^2)$.
Spin is ignored in our present schematic notation.
The vanishing of $b^2$ is a
necessary condition for CT to occur.
The quantity $b^2(Q^2)$ is closely related to the ratio of
the $T_S G_0$ and 1 terms of eq.\rep{eq:matel}, which vanishes if CT occurs.
But the vanishing of $b^2(Q^2)$  is not
sufficient for CT to occur as the energy must be high
enough for  the
escape time to be small  compared to the expansion time, see Sect.
{}~\ref{subs:td}.

In non-relativistic treatments the operator $T_H$ has the simple form
$T_H=\sum_i e_i e^{i \vec {q}\cdot \vec{r}_i}$.
Then \cite{FMS92,FMS931} for $q\bar{q}$
 systems
$b^2(Q^2)$ is given by
\bea
b^2(Q^2)=-16 {d\ln F(Q^2)\over dQ^2}. \label{eq:2body}
\eea
Thus $b^2(Q^2)$ is a constant for
oscillator model wave functions, but falls as $1/Q^2$ for models
in which the form factor falls as a power law such as for the
Coulomb
interaction.
Both a confining force and a one gluon exchange Coulomb force are important
features of quark models.  If one considers an interaction which is a sum
of both terms, one finds that $b^2(Q^2)\sim 1/Q^2$
for values of Q that are greater than the inverse
size of the system.

Relativistic versions of eq.~\rep{eq:2body} can be obtained using
light cone quantum mechanics. The results are qualitatively similar to the
non-relativistic case\cite{FMS92,FMS931}.
 Coulomb wave functions yield small values of
$b^2(Q^2)$, while oscillator wave functions lead to a $b^2(Q^2)$ which
rises slowly as Q$^2$ increases.
The above examples indicate that the presence of strong non-confining
potentials
which are singular at small distances
such as $1/r$ leads to
small values of $b^2 (Q^2)$, for not too high values of $Q^2$.

In modern constituent quark models (e.g. Refs. \cite{34.}-\cite{36.},
quark-quark short-ranged interactions are very important in determining the
energy and wave function. This could also be true in QCD \cite{shuang}.
As an example, we compute $b^2(Q^2)$
using the wave function of Capstick and Isgur \cite{34.}.
As shown in Fig. ~8
$b^2(Q^2)$ decreases significantly as Q$^2$
increases.

In $QCD$ the nucleon is expected to have many
Fock space components. One way to partially
account for these effects is to include a pion cloud or, as in the
Skyrmion model\cite{skyrme}, to treat
the entire nucleon as a soliton of pion fields.
The role of the $PLC$ in the Skyrme model was studied in Ref.
\cite{FMS92,FMS931} with the result, shown in
Fig. 8, that b$^2(Q^2)$ decreases noticeably with Q$^2$.
 The sharp bag surface prevents the same result from occuring
in the Stony Brook group's Refs.~\cite{33.},
\cite{40.}  modification of including a spherical
bag of free quarks at the center of the soliton.

In the cloudy bag model \cite{43.,44.} the pion
is a quantum fluctuation treated in perturbation theory. The results
\cite{FMS92,FMS931} using
the Regensburg version of the cloudy bag \cite{45.} are that
one sees a decrease of b$^2(Q^2$) due to  the usually significant
pion tail becoming  inessential as $Q^2$ increases towards 1 GeV$^2$. This
interesting effect can be studied in the domain of traditional nuclear
physics; see Sects. \rep{subs:search} and \rep{sec:summary}.

The final model we consider is the Li-Sterman  perturbative
treatment of the pion form factor.
The numerical results are shown in Fig.~8.
At high $Q^2$,
$b(Q^2)$ falls off more quickly than ${1\over Q}$ so that
including Sudakov effects causes the size of the wavepacket to be even
smaller than that expected from naive perturbative QCD.

The ``harmonic oscillator" and ``hydrogen atom" wave functions and the
Capstick-Isgur model are three examples of ``soft wave functions" in the
sense of Isgur and Llewellyn-Smith \cite{isgurls}: no explicit gluons are
present. The results shown in Fig.~8  indicate that a soft  wave function
can yield color transparency. The validity of pQCD is not a requirement for
the existence of color transparency.

\subsection{Time development}\label {subs:td}

Suppose a $PLC$ is produced in the interior of the nucleus. Any non-
eigenstate  undergoes time development. Here expansion occurs because the
starting system is defined to be small, see Sect.~\ref{sec:intro}.
This
expansion has been found to be a vital effect
\cite{FLFS88,jm90,jm91,kopel,ob,nnn93} for intermediate energies, $P_{lab}$
less than about 20 GeV/c.

We are concerned here  with time development in
nuclear quasielastic reactions. Thus
the (e,e'p) reaction provides an example,  with  straightforward
generalizations to other quasielastic
processes.
The amplitude for removing a proton from the shell model orbital
$\alpha$, ${\cal M_\alpha}$ and detecting an outgoing proton of momentum
$\vec p$:
\bea
{\cal M_\alpha} = \langle N, \vec {p}
 | ( 1 + T_S G_0 ) T_H |N,\alpha \rangle,\label{eq:matel}
\eea
where $T_H$ is the operator making the hard scattering,
 and the soft ejectile-nucleus
scattering amplitude $T_S$ is obtained by solving the
Lippman-Schwinger equation
$T_S= U + U G_0 T_S $ where U represents the interaction between the
ejected object and the nucleus,
and $G_0$ is the non-interacting Green's function.

The old fashioned approach is to treat the ejectile as proton. Then the
operator U is the optical potential U$^{opt}$, with
$U^{opt}=-i \sigma \rho(R)$,
where $\sigma$ is the proton-nucleon  total cross section, and $\rho(R)$ is
the nuclear density and R is the distance from the nuclear center. If the
proton wave function is computed from
$U^{opt}$, the proton wave is said to be
distorted (from the plane wave approximation). The use of
such a wave function in computing
 ${\cal M_\alpha}$ is
the distorted wave impulse approximation DWIA, where the ``impulse"
refers to the use of the free nucleon-nucleon cross section.
For simplicity, the
finite  range of the nucleon-nucleon interaction is
neglected here.
Only the imaginary part of U is kept because the real part of the NN
forward scattering amplitude is small at high energies.

But if the ejected object is a PLC, using
U$^{opt}$ is not appropriate. On the other hand, the ejectile expands as it
moves through the nucleus, so that one can not simply
neglect $T_S$. The need to include
this expansion
was recognized by Farrar et al.\cite{FLFS88} who
argued that the square of the transverse size is roughly
proportional to  the
distance travelled
$Z$ from the point of hard interaction where the PLC is formed. Thus the
``$\sigma$" that appears in the optical potential is replaced by one
that grows as the ejectile moves in the Z-direction
\bea
\sigma^{PLC}(Z) =(\sigma_{hard} + {Z\over l_c}[\sigma
-\sigma_{hard}])
\theta(l_c- Z) +\sigma\theta\left(Z-l_c\right). \label{eq:sigdif}
\eea
This equation is justified for hard stage of time development
in the leading logarithmic
approximation when perturbative QCD can
be applied \cite{FLFS88,BM88,FS88,DKMT}.
One can expect that
Eq.~\ref{eq:sigdif}
smoothly  interpolates between the hard and soft regimes.
A sudden change of $\sigma^{PLC}$ would be inconsistent with
the observation of an early (relatively low Q$^2$)
Bjorken scaling \cite{FS88}.

The time development of the $PLC$ can also be
obtained by modeling the ejectile-nucleus interaction as
$\hat U = - i \sigma (b^2) \rho (R)$,
recall Eq.~\rep{eq:sigpos}
Then
one can assume a baryonic basis and compute the relevant matrix elements
of $\sigma (b^2)$
\cite{jm90,jm91,kopel}.
Jennings and Miller \cite{jm90,jm91}
suggested that the Lippman-Schwinger equation
$T_S=\hat U (1 +G_0 T_S)$
could be solved by computing the term of first order in $\hat U$
and exponentiating the result. In this case,
U$^{opt}$
is modified by replacing the $\sigma$ by a cross section
$\sigma_{eff}$
given by,
\bea
\sigma_{eff} (Z) = \sigma \left(1 - e ^{- i \Delta p Z}\right)
\label{eq:ho}
\eea
where $\Delta p = {M^2_1 - M^2 \over 2 p} $. The eq. \rep{eq:ho} is
obtained by using
$\sigma(b^2)\propto b^2$ and a harmonic oscillator basis for
the soft features of the baryonic basis. (The hard features
are included in the operator $T_H$. )
Using the frozen approximation
of $p \rightarrow \infty$ leads to $\sigma_{eff} = 0$ and, color transparency
occurs. On the other hand sending $M_1$ to infinity
causes the old fashioned optical potential to emerge.
Greenberg and Miller
\cite{greenberg} showed generally that exponentiation
is a very  accurate approximation.

A more elaborate approach was taken later \cite{jm92} by using
measured matrix elements for deep inelastic scattering DIS
$\langle \alpha M^2_X | T_H(Q^2) \, | N \rangle$ and diffractive
dissociation DD $\langle N (\vec q) | 4 \pi \, Im \, \hat f
| \alpha, M^2_X \rangle$.
Here one gets contributions from all values of
M$_X^2$, so   that some CT effects appear at lower values of p, but
a complete cancellation of the interaction requires higher values of p
than the model of eq. \rep{eq:ho}. For small Z, the $\sigma_{eff}$
is approximately linear in Z as is
$\sigma^{PLC}$ of eq.~\rep{eq:sigdif}.
 Note also that the DD amplitudes
 used in obtaining $f$, lead to reasonable inelastic shadowing corrections
to total cross sections, recall Sect.\rep{subs:screening}.

Still another approach \cite{jm90,FS91,FGMS92,boffi,kj93} involves the
treating the baryon-nucleon amplitude in terms of a
finite number of baryonic states as
used to   describe the propagation of hadrons through
nuclei, see e.g.  \cite{Vanhove,Gottfried,coneschi,oldboris,Fialkovki,hf92}.
 To be brief, consider a two-state treatment (at least three are needed
to be consistent with data on cross section fluctuations \cite{FGMS92})
in which one  uses a
a matrix representation:
for the baryon-nucleon transition matrix T.
Then $\hat U =-i T \rho$ and
\bea
T=-i\sigma \pmatrix {
1 & - {\alpha \over \beta}   \cr \cr
-{\alpha \over \beta}  & \alpha^2/\beta^2 \cr }
.\eea
The eigenvalues of this matrix are an example of the Good-Walker diffractive
eigenvalues discussed in Sect. \ref{sec:soft}
The state with eigenvalue 0 has no interactions and is therefore
identified as the PLC
$|PLC \rangle = \alpha |N> + \beta |N^* \rangle$.
The PLC is expands with a time scale defined
by the difference in the square of the masses of the two baryonic states.
But
the absence of interactions required for complete
color transparency can only be
obtained if the T-matrix have at least one state of eigenvalue 0.
But several different papers\cite {oldboris,boffi,kj93}
use two state models without satisfying this
condition.  One general limitation of
any hadronic basis model
is the need to model the important states in  both
DIS and DD.
For large mass states the assumption that similar states are important
is in contradiction with QCD, with the observation in DIS but not in DD
of correlations between quantum numbers of particles in
opposite fragmentation regions.  This effect was modeled in
\cite{jm92} by introducing a cutoff at large M, but the value of the
overlapping integral for the  resonance region requires further investigation.

\subsection{Relevant data}
\label{subs:search}
It is natural to consider the (e,e'p) and (p,pp) processes for color
transparency searches.
The first published experiment aimed at
observing the effects of color transparency was the BNL $(p,pp)$
work of Carroll {\it et al.} \cite{bnl88}. The only other in the literature
is the SLAC (e,e'p) NE18 experiment \cite{ne18}.
We discuss each and the related theory issues in some detail.

\subsubsection{The BNL (p,pp) experiment}
The
beam momenta $p_L$ were 6, 10 and  12 GeV/c.
The experimental setup used was that for  proton hydrogen
 elastic scattering
at a center of mass angle of
90$^\circ$\cite{kirs}.
For the hydrogen target, the identification of an elastic scattering event
required only detecting  the outgoing
momentum of one proton and the angle of the other.
This is not sufficient for nuclear scattering because of the influence of
the
Fermi motion of the bound proton.
However, information from veto counters was used to suppress
inelastic events.

The data were originally plotted with an effective beam momentum,
$P_{eff}$
related to the invariant mass of the two outgoing  nucleons. We display
the
data vs.  a more intuitive quantity -$k_z$, the component of the momentum
of the struck nucleon calculated in the plane wave impulse
approximation.  The z-axis is defined by the beam direction.
The DWIA describes similar data at intermediate energies of $E_p= 1 GeV$
with accuracy of better than $20\%$ , see \cite{Belost} for review.
However the BNL data
are considerably above the DWIA
result of Refs.  \cite{bnl88,FLFS88,jm91,56.,FSZ93}- see the dot-dashed
curves in Figs. ~9  and 10 from \cite{FSZ93,jm93} where the
DWIA was applied to the
kinematics of \cite{bnl88}. The differences between the DWIA results
of the two figures arise from the use of different assumptions
about the experimental acceptance- the details are important.
In any case, the large
value of $d\sigma/d\sigma^B$ indicates
 the presence of a large transparency effect, but the
high energy drop caused considerable discussion\cite{bdet,rp88}.

There
is one technical point worth mentioning.
It is necessary to consider
whether or not a DWIA
 optical model calculation is relevant here.
Suppose an experiment measures a transition between the ground state and a
single well-defined final nuclear state. In that case,
the appropriate
calculation is a DWIA, with an optical potential obtained from $-i \sigma
\rho(R)$. The imaginary part of the optical potential accounts for
absorption from both pp inelastic scattering and p-A inelastic scattering,
with the later dominated by quasielastic knockout of a nucleon.
If the resulting energy loss is larger than the
experimental resolution, the quasielastic scattered particle is lost
and it is correct to use the optical potential, otherwise not.
The average energy loss in a
quasielastic scattering
is estimated as
1/(2$M_N \beta)\approx $ 70 MeV, where $\beta\approx  8 (GeV/c)^{-2}$
 determines the pN angular
distribution (e$^\beta t$),
and $M_N$ is the mass of the  nucleon. A proton
that scatters on its way out
is lost
if the experiment has resolution better than 70 MeV.
The energy resolution of the pioneering
BNL experiment was not known
exactly, but it was likely better than 70 MeV and it is probably acceptable
to use the optical
potential.
The  leaders of
the original experiment are carrying out new work with a new detector
$\cite{eva}$ which will have excellent
resolution, making the above discussion obsolete.

The color transparency
models which include expansion effects naturally produce an increase  of
the transparency consistent with the one observed in  \cite{bnl88} at
 6 and 10  GeV/c, see Fig.~9  and the dashed curves of Fig.~10.
 But they don't reproduce the drop in the
transparency indicated by the 12 GeV/c data.

Understanding the possible
nuclear results
at 12 GeV/c
requires some features of the
pp elastic scattering data. The energy dependence of the
90$^\circ$ angular distribution is of the form of 1/s$^{10}$ R(s)
where R(s) oscillates between 1 and 3 over the energy range of the BNL
experiment.
Ralston and Pire \cite{rp88}, suggested that the energy
dependence is caused by an interference between a hard amplitude which produces
a PLC, and a soft one
which involves a
large or blob-like
configuration BLC. The
Ralston and Pire idea is that
 the BLC is due to the Landshoff process\cite{rp88}.
Another mechanism
is that of Brodsky and de Teramond \cite{bdet}
 in which the two-baryon system couples to charmed quarks
(there is a small (6q) and a BLC which is a (6q,$c\bar c$) object)
The Brodsky-de Teramond idea originated from the observation
that the mass scale of the rapid energy
variation in
 $A_{NN}$\cite{kirs}  and in the measured transparency  matches that
of the charm threshold.  In both Refs.~\cite{rp88} and \cite{bdet}
the observed nuclear effect is claimed to
result from the suppression of the BLC in the nucleus: the observed ratio is
that of a pp cross section in the nucleus which varies smoothly with s
and the free pp cross section which has a bump.

However, the mechanisms are
imprecisely understood. For example,
the Landshoff mechanism  has a problem  in explaining the  significant
difference between pp and $p\bar p$ scattering cross
sections.
The energy dependence and phase of the Landshoff term are
different in pQCD from assumptions made in \cite{rp88},
see e.g. ref. \cite{Botts}.
Sudakov effects can be expected to supply
a set of configurations with a range of different  sizes.
Similarly even if
the details of the $c\bar c$ production amplitudes are not yet well
established,
threshold effects will naturally lead to mixtures of BLC and PLC.
Thus, it's natural to discuss high Q$^2$ elastic proton-proton scattering
in terms of configurations of different sizes.
Separating the contributing configurations into two, a PLC and a BLC
is only a simple first step.

Effects of the Fermi motion in treating the expansion process were
 discussed in the hadron basis approach in \cite{kj93,gardner,boffi}.
 The treatment was developed in the approximation where rescattering
amplitude behaves as a $\delta$-function of momentum transfer.
It was observed that in the discussed approximation transparency
substantially depends on $k_z$ at pre-asymptotic energies. The result was
that it became possible to construct a model\cite{jm93} which is
 able to describe BNL data at all energies.
The
solid curves of Fig.~10
 show the full calculation of Ref.\cite{jm93}
including the Ralston-Pire
intereference effect.
The $\sigma_{eff}$ of Ref.\cite{jm92} which leads to an increased
transparency at P$_L$=6 GeV/c compared with using
eq.\rep{eq:ho}.
The dashed curves ignore the RP effect.
The target-dependent and target
independent
uncertainties in the normalization of T are about 10\% and 25\%
respectively \cite{bnl88}.
 With these
experimental uncertainties in mind, the agreement between theory and
experiment is rather good.

We conclude that including color transparency effects is required to reproduce
the results of the BNL experiment.
This conclusion is supported by an entirely different
analysis due to Jain and Ralston \cite{jr93} and
by Anisovich et al.\cite{an92}.
We are unable to locate the
presence of time development effects in that analysis, and so
can not accept the detailed results of that paper.

\subsubsection{The SLAC (e,e'p) experiment}
\label{subs:slac}

If color transparency effects observed at BNL
are real they should be manifest in reactions
other  than (p,pp).
Thus the
recent measurement of the (e,e'p) reaction
 made at SLAC \cite{ne18} is very exciting.
 The NE18
collaboration measured cross sections for $^{12}$C, Fe and Ag targets
for momentum transfers Q$^2$ of 1, 3, 5 and 6.8 GeV$^2$. We quote results
presented in a recent preprint
\cite{ne18} discussing $^{12}$C. Data were taken over a range of
missing electron energy E$_m$ up to pion threshold and of missing transverse
momentum up to  p$_m = $
250 MeV/c. The longitudinal momentum p$_\parallel$ was limited to about
80 MeV/c.
The available data are shown in Fig.~11.
The transparency T(Q$^2$)
is defined in terms of the number of events N$_{data}$
measured in the region ${\cal R}$:
[-25 $ < E_m < $100 MeV; 0 $< p_m < $ 250 MeV/c and a plane wave impulse
approximation PWIA radiatively corrected model prediction for the same
quantity:
\bea
T(Q^2)\equiv 1.1 \pm 0.03  {\sum_{{\cal R}} N_{data}\over \sum_{{\cal R}}
N_{PWIA}}.
\label{eq:tdef}
\eea
The factor 1.1$\pm$ 0.03 accounts for correlation strength not contained
in the PWIA model, so it is assumed the radiative corrections are the same
for nucleons below and above the shell model Fermi sea.

The NE-18 experiment has made a significant achievement in
observing the
quasi-elastic (e,ep') reaction at Q$^2$ between 1 and 7 GeV$^2$.
 One is now faced with the
task of assessing
the data.
The quoted results of ref\cite{ne18}
are that no significant rise of the transparency with Q$^2$ is seen.
But how large can one reasonably expect CT effects to be for the SLAC
experiment?
One way to see is to
compare the data with the Glauber- DWIA
calculations.
Another way is to use models consistent with the BNL data
to compute CT effects for the (e,e'p) reaction.

First we discuss the
DWIA calculations of Frankfurt et al \cite{FSZ90},
Jennings and Miller \cite{jm91,jm92}, Benhar et al.\cite{ob},
 Frankfurt et al\cite{FSZ93},  and Miller\cite{gm}
and Kohama, Yazaki, and Seki \cite{kyz}.
The current SLAC experiment has good enough resolution for DWIA optical model
calculations to be relevant,  although
this conclusion depends on assuming the
accuracy of the unpublished radiative corrections.
This means that
the calculation of
Ref.\cite{kyz} , which
replaces the total cross section $\sigma$ in the
optical potential by the absorption cross
section, does not apply to the
the NE-18 $^{12}$C experiment.

Next we turn to specific results.
If one computes the transparency using the kinematic cuts of
Eq.~\rep{eq:tdef}, using nuclear wavefunctions obtained from the
Hartree Fock
approximation, uses
$\sigma = $  40 mb the result is $T(Q^2)=0.55$ for $^{12}$C as obtained by
Frankfurt et al \cite{FSZ93}.
For precise comparisons it is necessary to compute
T(Q$^2$) and  to use
good HF wave functions, otherwise errors of about 5\% for each effect
are induced.

But there is an interesting
energy dependence of $\sigma$  in the range of the SLAC experiment.
Values range from $\approx$
36 mb at Q$^2= 1 GeV^2$ (which effectively reduces to about 30 mb due
to fermi blocking which important for this low energy)
, to about
43 mb at Q$^2= 2 GeV^2$  and
40 mb at the highest
value of Q$^2$. This causes  the dip at Q$^2$ = 2 GeV$^2$ shown in Fig.~11.
This
effect has been confirmed by Greenberg \cite{wrgphd}.
The calculations of Greenberg and Miller\cite{gm,wrgphd}
include the vector nature of the photon in that
Dirac matrices $\gamma_\mu$ are evaluated between Dirac
bound and scattering states.
Greenberg's results are about
about  5 \% smaller than
the FSZ results,
due to
a different the radial form of the optical potential at $R\approx 0$.

Thus there is an unavoidable model dependence
of $\approx$  5\%, implying
that
detecting color transparency requires a
more substantial effect than 5 \%.

In any event, relevant DWIA calculations
must satisfy certain criteria:
(1)compute the relevant observable T(A) of eq.(\rep{eq:tdef})
(2) use nuclear wave functions which
reproduce the nuclear density
and spectral function
(3), include the energy dependence  of $\sigma$.
We show calculations satisfying  these criteria in Fig.~11.
As noted above the solid curve shows the Glauber DWIA. The dashed curve
shows the FSZ with eq.\rep{eq:sigdif} for the CT
while the dashed-dotted
curve uses a version of Greenberg, Jennings and Miller which is consistent
with the BNL (p,pp) data.
The effects of  nucleon-nucleon correlations produce a ``hole"
around the struck nucleon\cite{ob} and slow down the onset of color
transparency. Such effects are omitted in these results, so
the absolute value of all of the curves is slightly underestimated, but
the rate of increase with energy is noticeably overestimated for
the measured interval of $Q^2$.

 Both of the  calculations are
consistent with the BNL. There is no contradiction between the BNL data
and SLAC data.

Next we further discuss the effects of correlations
as emphasized by Benhar et al.\cite{ob}.
Although their calculations assume that
the experiment detects {\bf all} values of the missing momentum
$p_m$ and $p_\parallel$, leading to an
underestimate of about
 7\%, important
repulsive effects of nucleon-nucleon
correlations
in the optical potential are obtained.
This reduces final state
interactions and increases the computed transparency by
about 15 \% for a $^{12}$C target.
 Such an enhancement effect is apparently similar to CT, but
is independent of energy.
They also point out that presence of the hole delays the onset of CT since
initial phase of the expansion of the wave packet occurs in the region
 where no extra nucleons are present.
 Lee and Miller \cite{56.} found a similar result for the
(p,pp) reaction.
Correlations are present and must be included in a complete calculation.
However, the published calculations do not include the influence of the
energy dependence of the nuclear spectral function and relativistic
effects of the correlated nucleons.
Note that
the work of Refs.~\cite{kyz} and \cite{wambach}
find
 a small correlation effect,
(In Ref.~\cite{wambach} the  cross
section is integrated over all nucleon
momenta and excitation energies due
to elastic rescattering are  included,  so this work cannot be readily
compared with the  current data.)
Since CT in the discussed kinematics is
rather small,
 one of the pressing theoretical problems is to develop
methods to treat correlations beyond the optical
model approximation,  including initial and final state correlations in a
self-consistent way.

Yet another word of caution is necessary.   The EMC effect
\cite{EMC} tells us that the
structure function of a bound nucleon is smaller than that of a
free nucleon for $0.5<  x < \sim 0.8$.
But large x deep inelastic scattering is closely related to
the elastic form factor via the
Drell-Yan-West relation \cite{anytext}.
Thus a reduction
of the ratio of structure functions at
large x
should have an analog  in large Q$^2$ quasielastic reactions at $x=1$.
This effect should lead to a $Q^2$-dependent decrease of transparency
which  masks somewhat the onset of transparency \cite{FS88,FSZ93}.
 In particular, one of  predictions of color fluctuations for nuclear
 structure is
the suppression of PLC in bound nucleons \cite{FS85}, which
 leads to a decrease of the bound nucleon form factor. This effect rather
 strongly depends on the momentum of the struck nucleon and in the
 kinematics of NE-18
it may lead to depletion of transparency for carbon at large $Q^2$ by
 about  $7~\%$. This effect is included in the calculation of \cite{FSZ93}
 presented in Fig.~11.

\subsubsection{Rescattering conquers time development}

The practical problem in looking for CT effects in experiments at $Q^2$
from about one to a few GeV$^2$ is that the assumed $PLC$ expands rapidly
while propagating through the nucleus. To observe CT at
intermediate values of $Q^2$ it is necessary to suppress the effects of
wavepacket expansion.  This can be achieved by using the lightest nuclear
targets, where the propagation distances are small.  But then, the
transparency is close to unity, causing the effects of CT in $(e, e'p)$
reactions to be small.  However, if one studies a process where the produced
system can {\bf only} be produced by an interaction  in the final state, a
double scattering event, then the color coherent effects would manifest
themselves as a decrease of the probability for final-state interactions
with increasing $Q^2$.  One could then observe an effect decreasing from
the value expected without CT (Glauber-value) to zero.  Thus, the measured
cross section is to be compared with a vanishing quantity so that the
relevant ratio of cross sections runs from 1 to infinity.
The first calculations\cite{double} show that substantial CT effects
are observable in the
(e,e'pp) reactions on $^{4,3}He$ targets. Such experiments can be done at
the Continuous Electron Beam Accelerator Facility CEBAF.

Another idea involves pionic degrees of freedom. Sect.\ref{subs:small}
shows
that probing a nucleon at intermediate momentum transfers ($Q^2$ about a
few GeV$^2$) may produce a small system without a pion cloud.
This
cloud-stripping effect can be studied
by considering processes that require a pion
exchange to proceed. An example is the quasielastic production of the
$\Delta^{++}$ in electron scattering - the $(e,e^\prime\Delta^{++})$
reaction. The initial singly charged object is knocked out of the nucleus
by the virtual photon and converts to a a $\Delta^{++}$ by emitting or
absorbing a charged pion. But pionic coupling to small-sized systems is
suppressed, so  this cross section for quasielastic production of
$\Delta^{++}$'s should fall faster with increasing  $Q^2$ than the
predictions of conventional theories. Calculations are now in progress
\cite{FLMS94}.
This could be a new  kind of transparency that
involves pions, so the present authors have invoked
 the name ``chiral transparency"\cite{FMS92}.

A final comment concerns the use of the (e,e') reaction to study color
transparency effects\cite{vijay}. We find this problematic because the
measurement includes processes in which nucleon excited states are
produced and it is technically difficult to compute the effects of the
final state interactions. At first glance, these should be given by
-i$\sigma_{abs}\rho$, but this is not done in the calculations. Moreover,
the biggest effects are in the region where the relativistic motion of the
nucleon is important.

\section{Color fluctuations in nucleons  and nucleus-nucleus collisions}
\label{sec:heavyion}
Relativistic heavy ion collisions are considered
to be one of the most
promising
ways of studying
nuclear matter under extreme conditions. In order to establish
whether something unusual happens in these collisions
one has first to build realistic models of
these collisions incorporating more conventional phenomena.
 Since fluctuations of
interaction cross sections arising from color fluctuations constitute an
essential element of strong interaction dynamics, it is crucial to take them
into account in building realistic models of nucleus-nucleus collisions.
Moreover the very existence of color coherence phenomena
lead to significant  dependence of parton structure of colliding nuclei
on whether or not the collision is a central one.
This property is a promising
new handle to regulate
properties of the initial hadron state produced in AA collisions.

The energy range where these fluctuations
become important is dictated by the
coherence length (or freeze out  parameter)
for color fluctuations -  $2E_N/(m^2_X-m^2_N)\equiv\l_c$.
Coherence effects are expected to become noticeable
as soon as this distance
 exceeds 2-3 fm, the typical
distance between nucleons in
nuclei ($E_A/A \ge 10 GeV$, for a typical soft scale $m_x\approx 1.2 $ GeV).

 If we use interval duality between quark and hadron degrees of freedom
in dispersion sum rules as a guide to estimate  the scale of masses
for the developed color fluctuations $m_X$  should be
of the order 2-3 GeV. Thus
 for  energies  above
$E_A/A \ge 50A^{1\over 3} GeV$
the significant  parts
of the fluctuations are
inhibited, so that one may
treat the nucleons in nuclei  as frozen in their initial
configurations during the entire collision process.
The major effect coming from the
the presence of fluctuations is similar
to that for production of lepton pairs
in the propagation of the positronium through
media (Fig.~2 and Sect. 2). In positronium case it turned
out to be more likely to emit two  pairs in one
event than one would infer from the
independent emission picture. Similarly,
the number of nucleons involved in a  central
nucleus-nucleus collision should fluctuate with the
transverse size of quark-gluon configuration in the nucleon - see Fig.~1
Nucleons in a small size configuration
knock out much fewer nucleons
than average, but nucleons in a large size configuration knock out
much more.

    This  effect is implicit in string models of hadron-nucleus
interactions.  In these models the effective transverse size of a hadron, $d$,
is $\approx l\sin\theta$ where $l$ is the distance between the quark and
di-quark (or antiquark) in the hadron, and $\theta$ is the angle between the
momentum of the hadron and the direction of its string.  Since $d$ is frozen
during the collision with the
target nucleus, the number of struck nucleons, $N$,
should increase as $d^2$.  Thus $N$ fluctuates from collision to collision
even at the same nucleus impact parameter, a consequence of different string
orientations with respect to the reaction axis.  These fluctuations are in
addition to Poisson fluctuations, and, as we see below, quantitatively more
important.  The variation of cross-section with $d$ is neglected in all
current modeling of nucleus-nucleus collisions,
 where one uses the average over $d$ in the
interaction with one nucleon, rather than more correctly {\em first generating
collisions with strings of given $d$ and then averaging over events produced
in the collisions of projectiles with different $d$}.

\subsection{Color opacity effects}
To illustrate the magnitude of fluctuations in
AA collisions resulting from color
fluctuations of the cross section let us first consider
 fluctuations in the number of struck nucleons in interactions at fixed
impact parameter.  In particular, the scaled variance, $\omega_N =
(\av{N^2} -\av{N}^2)/\av{N}$, of the number of collisions, $N$, for a proton
propagating through a large nucleus is an interesting quantity.
  Poisson fluctuations in the number
of collisions would lead to $\omega_N =1$, but inclusion of color fluctuations
leads to a significant increase of the variance \cite{BBFHS91}:
\beq
\omega_N =1 + \av{N} \os,   \label{eq:omega}
\eeq
where $\os=(<\sigma^2>-<\sigma>^2)/<\sigma>^2$.
More realistically, nucleon-nucleon
correlations arising from both the Pauli principle and short range repulsions
decrease the variance because fluctuations involving
too many nucleons at one
impact parameter are suppressed. In the absence of color fluctuations
$\omega_N$ is reduced
to $1 - \alpha$, where $\alpha$ is the size of the nucleon correlation hole.
Nuclear calculations with realistic short-range correlations between nucleons
lead to \cite{Pi,BBFHS91} $\alpha \approx 0.37$.  Taking the two effects into
account we find that for central hA collisions,
\beq
  \omega_N=1-\alpha+\os (\av{N}-\alpha)  \, .\label{os}
\eeq
The fluctuations in nuclear AB collisions can be calculated analogously to
the pA case. For central collisions,
\beq
    \omega_N
    =1-\alpha-\beta+ \gamma+\os(N_{pA}+N_{pB}-\alpha-\beta)   , \label{osAB}
\eeq
where $\beta \approx \alpha$ is the analogous correlation parameter for
the projectile nucleus B, and $\gamma \la \alpha$ involves the product of both
correlation holes \cite{BBFHS91}.  For $B\ll A$, as in CERN experiments with
$^{16}$O or $^{32}$S projectiles, $N_{pA}=2{\bar\sigma}\rho_0 R_A$, and
$N_{pB}\simeq 3\s\rho_0 R_B/2\simeq$ 2-3.  Thus for heavy nuclear targets,
color fluctuations increase $\omega_N$ from 1.26 to $\sim$ 2.9-3.1 for $\os =
0.2$.

It is possible to confront this consequence
of color fluctuation
physics
with
the CERN nuclear beam data using an
analysis of the NA34 transverse energy $E_t$ distribution\cite{NA34}.
This analysis  demonstrated that
their
geometrical model
which for
multiple nucleon-nucleon
collisions
 reproduces
the multiplicity and transverse energy spectra
to good accuracy. It was, however, pointed
out by Baym et al. \cite{BFS}  that the fluctuations extracted from
experiment were much larger than those expected from a microscopic
model with fluctuations given by the known nucleon-nucleon
fluctuation in transverse energy.
   It turns out
that Eq.\rep{osAB} with a value of
$\os \sim 0.2$, (close to the one derived
from the analysis of diffraction data - section 4)
produces increase in the amount
of the
fluctuations close to the  one suggested by the
NA34 data - see Fig.~12.

Future studies will involve moments  higher than two.
    A useful practical prescription to implement such color fluctuations
in modeling AA collisions is based on the factorization
assumption for NN interactions
in different configurations (cf. discussion in
\cite{BBFHS91}).  To generate interaction cross sections for a pair of
interacting nucleons, labelled 1 and 2, one generates $\sigma$
 for each of the
nucleons with individual weights $P(\sigma_1)$ and $P(\sigma_2)$, and then
uses
a cross section $\sigma_1 \sigma_2/\av{\sigma}$.

    Furthermore,
the dependence of parton distributions on the size
of the nucleon configuration must be included in modelling
 fluctuations of hard parton-parton interactions.
This size dependence can be studied experimentally
in pA collisions see \cite{FS85}.
Since the increase of the effective transverse
size of a hadron allows emission
of softer partons, the simplest assumption is
that the in-medium parton distributions $p_N(x,Q^2,\sigma)$ scale as
\beq
 p_N(x,Q^2,\sigma) = p_N(x,Q^2 \sigma/ \av{\sigma}),
\label{close}
\eeq
where $p_N(x,Q^2)$ is the parton distribution of a free nucleon.
This is
similar to the rescaling model of the EMC effect \cite{Close}.

One should note that existing
models and codes include correlations between nucleons only in an approximate
way whose effect is to overestimate fluctuations in the number of collisions
originating from the fluctuations of the number of participating nucleons
helping these models to fit the tails of measured $E_t$ distributions
for more detailed discussion see
\cite{BBFHS91}.

It is worth emphasizing that including effects of color fluctuations in
the picture of inelastic nucleus - nucleus interactions solves a problem of
incorporating
the effects of inelastic shadowing which
are known to be present in high-energy interactions with nuclei, recall
 Sect. 4. It is possible
to demonstrate that though the procedure described here
resembles a classical
probabalistic
 description it adequately
reproduces the rules which follow from the quantum
field theory analysis (Heiselberg et al. in progress)  based
on the Reggeon calculus \cite{Gribov67} and  Abramovski$\breve{{\rm i}}$,
 Gribov, Kancheli cutting rules \cite{AGK}.

\subsection{Color fluctuations
and central collisions}
In nucleus-nucleus collisions
involving hard interactions of partons
 with small
$x$ one has
to include the $\sigma$ and impact parameter dependence of the parton
distributions. This gives significant
depletion-shadowing and enhancement effects.
\cite{FS90}. This is because medium modifications of the parton distributions
 are enhanced for central AA collisions.
One of expected effects is the strong suppression of $J/\Psi$
 production at RHIC
and LHC
energies, $\Upsilon$ production at
 LHC energies, etc. As a result $J/\Psi,\Upsilon-$meson
production at small x are hardly can be used as  the trigger
for new forms of nuclear matter.

Color fluctuations
allow one a
possibility of selecting rare configurations in colliding nuclei using high
$E_t$ triggers.  As shown in ref.~\cite {BF93}, by selecting AB collision
events with a given $E_t$, weighted by a factor
$E_t^n /\av{E_t}^n$,
one can strongly enhance the
probability of large interaction cross-section configurations in colliding
nuclei.  Note, for example, that the probability for a given nucleon to be in
the ``fat" configuration $\sigma /\av{\sigma}$ =1.25, for $n=1-4$
in the distribution
(8) is $\le$ 0.52-0.95 (and for $\sigma /\av{\sigma}$ =1.5, $\le$ 0.30-0.80).
As a
result, transitions to percolated configurations \cite{GB78}, where ``fat"
nucleons form a connected net in the nucleus, are possible.  Parton
distributions in such percolated nets should be significantly different from
inclusive parton distributions [cf. eq.  (\rep{close})]; therefore we expect
that parton distributions measured with a high $E_t$ trigger should be
significantly different from parton distributions
in a free nucleon
\cite{BF93},
with
the weight
of partons with large $x$
suppressed with increase of the
``fatness" of a nucleon.
Thus investigation of color fluctuations may help to built effective
trigger for the change of collective properties of hadron matter .

\subsection{Leading hadron transparency effects in the central AA collisions}

  Finally, we mention that color fluctuations into small-size configurations
 lead to interesting color transparency phenomena in the projectile
fragmentation region for $E_A/A \ge$ 100 GeV, similar to the decrease in
absorption of ultrarelativistic positronium discussed in
Sect.~\rep{sec:qed}.
Indeed, if a nucleonic PLC enters the nucleus
it can fly through it
practically without interaction. Since all large
size configurations would be
filtered out,
the PLC escaping from the nucleus would decay both
into nucleons and excited states.

One can express  the probability for a
given central nucleon of the projectile to
go through the center of a heavy nucleus in a central AA collision -
$ W^{N\rightarrow N}$ without any interactions through $P(\sigma)$ as
\beq
 W^{N\rightarrow N}= \left[ \int exp[-\sigma T(B)/2]
P(\sigma) d \sigma\right] ^2 \label{WNN}
\eeq
where $T(B)= \sigma_{tot}(NN) \int dz \rho_A(B,z)$ and
$\rho_A(B,z)$ is nuclear
density.  Eq. \rep{WNN} leads to
significantly higher probability of penetration than the
classical probabalistic expression \rep{expabs}
For realistic $P(\sigma)$ and $A \sim 240$ the probability
is expected
to be on the order of 1$\%$ as compared to
naive expectation of $\sim 2\cdot 10^{-4}$.

The total probability for a nucleon to
transform into  diffractive  states
(including nucleon),$ W^{N\rightarrow X}_{tot}$
can also be expressed through $P(\sigma)$ as

\beq
 W^{N\rightarrow X}_{tot}=
\int exp[-\sigma T(B)] P(\sigma) d \sigma \label{WNA}
\eeq

One can see that
$ W^{N\rightarrow X}_{tot} \gg  W^{N\rightarrow N}$
in the limit $A \rightarrow \infty $, recall Eq.~\rep{ptop}.
For
$A \sim 240$  probability for the nucleon to
fragment into a diffractive
system \cite{FS911} is
$ W^{N\rightarrow X}_{tot}-  W^{N\rightarrow N} \sim 1-2\%$.

There are several possibilities to look for these effects experimentally.
One is to consider a collision of a light nucleus, like $^4He$ on heavy nuclei,
 use a high $E_t$ trigger to
select  collisions at small impact
parameters, and investigate the yield  of leading nucleon spectators
($\alpha_N \equiv A p_N/ p_A \ge 1, p_t \sim 0)$. Another
possibility is  to use the Drell -Yan trigger ($\mu^+-\mu^-$ production)
to enhance
central collisions \cite{FS85}.
 One can also try
to look for production of leading nucleon
resonances at $\alpha_{N^*} >1, p_t \sim 0$.
 Though   the   measurement   of   the
production of nucleon resonances in  the  central  nucleus-nucleus
collisions seems to be  a much more difficult task than studying
nucleon production,  it  is  not  hopeless  because  the production  of
leading  pions  in central  nucleus-nucleus  collisions   is
strongly suppressed.

\section{Outlook}
\label{sec:summary}

It is often perceived that  only
quasielastic
reactions
$p(e)+A\rightarrow p(e)\prime
+N + (A-1)^*$
are
worth using
for studying color transparency CT.
Though this is the simplest reaction to measure, a much wider range of
reactions would be necessary to build a sufficiently complete picture.

\noindent
i) Use of different projectiles is necessary, since different hadrons
have different PLC probabilities. Furthermore, the relative
importance of various
reaction
mechanisms, such as that of  Landshoff, will vary with the projectile.

\noindent
ii) It is necessary to study CT in quasielastic processes
off nuclei
in the limit of
large energies ($\sim 500 GeV$ and large ($\sim 10 GeV^2$)
but fixed momentum transfer, -t.
In this limit the PLC expansion is not important for the projectile and the
leading hadron and the cross section
may vary as A$^{2/3}$ instead of A$^{1/3}$. Furthermore, the interplay of
contributions of large and small distances could
be different from that occuring
in  large angle scattering reactions at lower energies.

\noindent iii)
One important consequence of the CT picture is that cross sections
for production of diffractive states in
large energy wide
angle reactions
\cite{jm90,FS91} are large.
So it is important to look for the processes like
$p+A\rightarrow N^{*} (N\pi, N\pi\pi) +N$. Knowledge of the
relative abundance of
resonances and continuum would help in building more realistic models for the
expansion of PLC.

\noindent
iv) Experimental studies of two-body large angle reactions
\cite{Car2} have demonstrated that reactions where quark exchange is
allowed have substantially larger cross sections than other reactions.
 Since the quark-exchange
diagrams are expected to be dominated by small distances, the  study of the
reactions like   $p+A\rightarrow \Delta^{o} +p + (A-1)^*$  would be
extremely important.

\noindent
v) To obtain a more detailed knowledge of the interaction of PLC with
nuclei it is worth looking at processes where projectile could interact
second time  during the passage of  the nucleus \cite{FS91,double}.
This can be
done by studying recoil nucleons with momenta $k \geq 300 MeV/c$. The
number of such nucleons should decrease substantially with onset  of CT.
An
important advantage of this reaction is
that the effect of the disappearance of the
final state interaction can be studied using the lightest
nuclei (${^3}He,
 {^4}He$).
 Another advantage is that internucleon distances probed are
not large- 1-2 Fm so
effects of  the wave packet expansion become
less important. Thus
one may search for CT effects
at kinematics  where CT for the  $p+A\rightarrow p\prime +N$ process
is too small to be reliably observed.

\noindent
vi) Detailed studies of soft coherent nuclear diffraction for a
wide range of A
is necessary to map the cross section fluctuation distributions
of various hadrons.

\noindent
vii) Studies of hard diffraction using lepton and hadron beams
are necessary to  investigate the transition from soft Pomeron physics
to perturbative hard physics.

\noindent
viii) Studies of hadron interactions with nuclei with a hard trigger
(like a Drell-Yan pair) are necessary to study correlations between parton
distributions and the transverse size of quark-gluon configurations in hadrons
\cite{FS85}.

Obviously, performing this program will require experiments at a wide
range of experimental facilities including CEBAF, BNL (AGS and RHIC), FNAL
(fixed target program), CERN (muon beam
and heavy ions), DESY (HERA, and HERMIS).

We thank the USDOE and the US-Israel   BSF for partial financial support.
We are especially indebted to our co-authors who have made this review
possible:
G. A. Baym, B. Blattel,  A. Bulgac,
S.J. Brodsky,  K.S. Egiyan, G.R.Farrar
W.~R.~Greenberg, J. F. Gunion, H. Heiselberg,  B.~K.~Jennings,
T.-S.H. Lee, S. Liuti, A. H. Mueller,
M. M. Sargsyan, and  M. Zhalov. We thank L. Sorensen for his comments on
the introduction.

\vfill\eject

\vfill\eject

\noindent
Figure Captions

\noindent
1. Hadronic size fluctuations in collisions with nuclei. B is the impact
parameter. Projectile configurations of larger size interact with more
target nucleons.
\vskip0.1truein

\noindent
2. Production of two l $\bar l$ pairs. Two different target atoms,
represented by the upper and lower horizontal lines,  are involved
\vskip0.1truein

\noindent
3. Photon-gluon fusion model.  The incident
virtual photon, $\gamma^*$ breaks up into a $q \bar q$ pair before hitting
the target. The pair interacts with the target gluons (dashed lines).
\vskip0.1truein

\noindent
4. Typical two-gluon exchange contribution to $\gamma^*(q) p\rightarrow
V p$. The target gluons are represented by wavy lines. The final
four-momentum of the vector meson, V, is  $ q + \Delta$, and the target
proton acquires a four-momentum $-\Delta$.
\vskip0.1truein

\noindent
5. Illustrative diagram for hard diffractive production of jets when an
incident pion breaks into a $q\bar q$ pair before hitting the target.
The target gluons are represented by wavy lines.
\vskip0.1truein

\noindent
6. Cross section probability
for pions  P$_\pi(\sigma)$ and nucleons P$_N(\sigma)$ as extracted from
experimental data.  P$_\pi(\sigma=0)$ is compared with the
perturbative QCD prediction.
\vskip0.1truein

\noindent
7. $\sigma_{diff}(A)$ for pion and proton beams. The pionic data are from
Ref. 92, and  the nucleon data from Ref. 91.
The spread in the calculations results from the use of different
 P$(\sigma)$ with the same dispersion of cross section, $\omega$.
 The A-dependence is more
rapid than $A^{1/3}$, confirming the presence of cross section fluctuations
near the average.
\vskip0.1truein

\noindent
8. $b^2(Q^2)$ of Equation (5.1) for various models discussed in the text.
The rapid fall  shows that many models predict the formation of
PLC at moderate values of $Q^2$.
\vskip0.1truein

\noindent
9. (p,pp) calculations of Ref. 4 compared with the data of Ref. 130.
for three different beam energies and four values of $k_z$,
the momentum of the struck nucleon defined in the text.
The qualitative agreement between the predictions of color transparency
(solid curves labelled CT)
and the
data, along with the small values of the DWIA calculations (dashed-dot curves)
hints at a significant effect of color transparency.
\vskip0.1truein

\noindent
10. (p,pp) calculations of Ref. 137  compared with the data of Ref. 130
for three different beam energies and four values of $k_z$,
the momentum of the struck nucleon defined in the text.
The qualitative agreement between the predictions of color transparency
(solid curves labelled CT including the effects of larger sized
configurations according to Ralston-Pire;  dashed curves
without those effects)
along with the small values of the DWIA calculations (dashed-dot curves)
hints at a significant effect of color transparency.
\vskip0.1truein

\noindent
11. NE18 data [15] compared with calculations of Ref. 4
using $M^2 -m^2$ of Eq. (1.1)
as $\Delta M^2 = 0.7 GeV^2$. The curve labelled GJM93
use the $\sigma_{eff}$ of Ref. 143 and the proton-nucleon cross sections of
Refs. 145 and 146.
\vskip0.1truein

\noindent
12. Dispersion of transverse energy in heavy ion collisions.
The data are from Ref. 155. The agreement between the theory and the data
indicate the presence of configurations with larger than average cross
sections.

 \end{document}